\documentclass[notref,notcite,fleqn,12pt]{article}
\usepackage{amsmath,epic,graphicx,amssymb,times}

\makeatletter

\@addtoreset{equation}{section}
\makeatother

\newcommand{\bmath}{\boldmath}
\newcommand{\ubmath}{\unboldmath}

\newcommand{\be}{\begin{eqnarray}}
\newcommand{\ee}{\end{eqnarray}}
\newcommand{\bes}[1]{\begin{subequations}\label{#1}\begin{eqnarray}}
\newcommand{\ees}{\end{eqnarray}\end{subequations}}
\newcommand{\nn}{\nonumber}
\newcommand{\lrra}{\longleftrightarrow}

\newcommand{\cpd}[2]{{\stackrel{(#1)}{P_{\ }}}_{\!\!#2}}
\newcommand{\cjd}[2]{{\stackrel{(#1)}{J_{\ }}}{}^{\!#2}}
\newcommand{\cjo}[2]{{\stackrel{(#1)}{J_{\ }}}{}_{\!#2}}

\newcommand{\cqd}[2]{{\stackrel{(#1)}{Q_{\ }}}_{\!\!#2}}
\newcommand{\cld}[2]{{\stackrel{(#1)}{\Lambda_{\ }}}_{\!\!#2}}

\newcommand{\cpt}[2]{P_{#2}^{(#1)}}
\newcommand{\cqt}[2]{Q_{#2}^{(#1)}}
\newcommand{\clt}[2]{\Lambda_{#2}^{(#1)}}

\newcommand{\cCt}[2]{{\mathcal{C}}^{(#1)}_{#2}}
\newcommand{\cCct}[2]{{{\mathfrak{L}}^{(#1)#2}}}
\newcommand{\cCd}[2]{{\stackrel{(#1)}{\mathcal{C}_{\ }}}_{\!\!#2}}
\newcommand{\cCcd}[2]{{\stackrel{(#1)}{\mathfrak{L}_{\ }}}{}^{#2}}
\newcommand{\cCit}[2]{{{\mathfrak{C}}^{(#1)#2}}}
\newcommand{\cCid}[2]{{\stackrel{(#1)}{\mathfrak{C}_{\ }}}{}^{#2}}

\newcommand{\E}{E_{10}}

\newcommand{\K}{K(E_{10})}

\newcommand{\reals}{\mathbb{R}}
\newcommand{\ints}{\mathbb{Z}}
\newcommand{\mf}{\mathfrak}

\newcommand{\ra}{\rightarrow}

\newcommand{\p}{\partial}
\renewcommand{\a}{\alpha}
\renewcommand{\d}{\delta}

\newcommand{\eps}{\epsilon}
\renewcommand{\o}{\omega}
\newcommand{\s}{\sigma}

\renewcommand{\L}{\Lambda}

\newcommand{\G}{\Gamma}
\renewcommand{\O}{\Omega}
\newcommand{\tO}{\tilde{\Omega}}

\newcommand{\cV}{{\mathcal{V}}}
\newcommand{\cC}{{\mathcal{C}}}
\newcommand{\cQ}{{\mathcal{Q}}}
\newcommand{\cP}{{\mathcal{P}}}
\newcommand{\cD}{{\mathcal{D}}}

\newcommand{\cG}{{\mathcal{G}}}
\newcommand{\cM}{{\mathcal{M}}}
\newcommand{\cDso}{D}
\newcommand{\cJ}{{\mathcal{J}}}
\newcommand{\cL}{{\mathcal{L}}}

\newcommand{\cS}{{\mathfrak{S}}}

\newcommand{\op}{$\!\!\!\!\oplus\!\!\!\!$}

\allowdisplaybreaks

\begin{document}

{\flushright IHES/P/07/30\\AEI-2007-138\\[15mm]}

\begin{center}
{\Large \bf
Constraints and the \bmath$E_{10}\,$\ubmath Coset Model}\\[1cm]
Thibault Damour\footnotemark[1], Axel Kleinschmidt\footnotemark[2] and
  Hermann Nicolai\footnotemark[2]\\[5mm]
\footnotemark[1]{\sl  Institut des Hautes Etudes Scientifiques\\
     35, Route de Chartres, F-91440 Bures-sur-Yvette, France}\\[3mm]
\footnotemark[2]{\sl  Max-Planck-Institut f\"ur Gravitationsphysik,
  Albert-Einstein-Institut \\
     M\"uhlenberg 1, D-14476 Potsdam, Germany} \\[7mm]
\begin{tabular}{p{12cm}}
\hspace{5mm}{\footnotesize {\bf Abstract:} We continue the study of the
one-dimensional $E_{10}$ coset model (massless spinning particle motion
 on $E_{10}/K(E_{10})$) whose
dynamics at low levels is known to coincide with the equations of motion
of maximal supergravity theories in appropriate truncations. We show that
the coset dynamics (truncated at levels $\ell \leq 3$) can be
consistently restricted by requiring the vanishing of a set of
constraints which are in one-to-one correspondence with the canonical
constraints of supergravity. Hence, the resulting constrained $\s$-model
dynamics captures the full (constrained) supergravity dynamics in this
truncation. Remarkably, the bosonic constraints are found to be
expressible in a Sugawara-like (current $\!\times\!$ current) form in
terms of the conserved $E_{10}$ Noether current, and transform
covariantly under an upper parabolic subgroup $\E^{+}\subset\E$.
We discuss the possible implications of this result, and in particular
exhibit a tantalising link with the usual affine Sugawara construction
in the truncation of $E_{10}$ to its affine subgroup $E_9$.}
\end{tabular}\\[5mm]
\end{center}

\newpage

\begin{section}{Introduction}

Work on the symmetry structure of maximal supergravity
theories has revealed a remarkable link between geodesic motion
of a massless spinning particle on an $E_{10}/K(E_{10})$ coset
manifold and the dynamics of maximal supergravity theories
\cite{Damour:2002cu,Kleinschmidt:2004dy,Damour:2004zy,Kleinschmidt:2004rg,
  Damour:2005zs,de
  Buyl:2005mt,Kleinschmidt:2006tm,Damour:2006xu}. In contrast to an
earlier proposal \cite{West:2000ga,West:2001as,West:2003fc} aiming for an
11-dimensional covariant formulation of M theory exhibiting
$E_{11}$ invariance, the one-dimensional $E_{10}$ coset model
corresponds, on the supergravity side, to a (10 + 1)-dimensional
gauge-fixed formulation of the supergravity dynamics, as it arises
in studies of the near space-like singularity limit
\cite{Damour:2000wm,Damour:2000hv,Damour:2002et}. The
reformulation of the dynamics as a `cosmological billiard'
facilitates a systematic dynamical treatment, and directly
motivates the conjecture \cite{Damour:2002cu} that M theory
is (holographically) equivalent to a
`one-dimensional' non-linear $\sigma$-model living on the
infinite-dimensional coset manifold $E_{10}/K(E_{10})$.
Ref. \cite{Damour:2002cu} showed that the null geodesic
motion on $E_{10}/K(E_{10})$, when truncated
to low levels, is equivalent to a truncated version of the bosonic
dynamical equations of  maximal supergravity  where
only first order spatial gradients are retained. This equivalence
was extended by including the fermions (neglecting spatial
gradients) in Refs.\cite{Damour:2005zs,de Buyl:2005mt,Damour:2006xu}.
Some further evidence for a correspondence between M theory and the
$E_{10}$ coset model came from relating  M theory one-loop corrections
to certain high-level contributions to the coset action
\cite{Damour:2005zb}.

As is well known, in a canonical treatment of gravity
and supergravity, where space-time is foliated into a sequence of
spacelike hypersurfaces, the dynamical equations have to be
supplemented by {\em constraint equations} (to be imposed on the
initial data). For instance, in the case of pure gravity
these are the Hamiltonian and diffeomorphism constraints.
In the present contribution we study how such constraint
equations, which are necessary for recovering the full supergravity
system, can be consistently incorporated into the coset model
approach of \cite{Damour:2002cu}. As formulated there, this model
already incorporates (a close analog of) the Hamiltonian
constraint in the form of a null-motion constraint expressing
reparametrisation invariance of the worldline. We shall
therefore focus here on the other constraints, and study
to which extent they are compatible with the Kac--Moody symmetry
structure of these models (not manifest in the standard Hamiltonian
formulation of gravity). The consistency of the usual supergravity
constraints with the dynamical equations in the context of homogeneous
cosmological solutions was already studied long ago \cite{DHHS}.
Here, we are interested in establishing, {\em purely within the
context of the $E_{10}/K(E_{10})$ coset model}, the consistency of
requiring the vanishing of certain {\em bilinear} quantities in the
coset variables, either quadratic in the coset velocities $\cP$ (for
the bosonic constraints $\cC$), or consisting of a product of $\cP$
and the fermionic gravitino variables $\psi$ (for the supersymmetry
constraint $\cS$). Namely, we shall show that in the same consistent
truncation employed for the dynamical equations, one can define
bosonic and fermionic constraints of this type (on the massless
spinning particle) which are {\em weakly conserved}\footnote{We use
  the word `weakly' in the (constrained dynamics) sense of `modulo the
  constraints'. In other words, a set of constraints $\cC$ is weakly
  conserved iff $d \cC/dt $ vanishes {\em modulo $\cC$}.} along the
coset motion, thereby defining a constraint surface in the coset phase
space preserved by the geodesic motion. We will spell out the details
of this result only for $D=11$ supergravity
\cite{CJS}, but have no doubt that it carries over to the other maximal
and non-maximal cases (some of the supergravity constraint equations
rewritten in coset variables were already given in
\cite{Damour:2006xu,Hillmann:2006ic}). In this way {\em all} $D=11$
supergravity equations have been accommodated within the $E_{10}$
model.

In addition to the weak conservation of the constraints we find that
the equations describing the time evolution of the constraints exhibit
a triangular structure reminiscent of a highest weight representation,
cf.~(\ref{timederconst}). Studying the tensor structure of the
relevant constraints reveals two further structures, namely:
\begin{itemize}
\item One can redefine the bosonic constraints $\cC$ into an equivalent
 set $\mathfrak{L}$ of explicitly time-independent (hence {\em strongly
 conserved}) `Sugawara-like' expressions bilinear in the conserved
 Noether current (or charge) $\cJ$ associated to the rigid $E_{10}$
 symmetry of the $E_{10}/K(E_{10})$ coset action.
\item At least for the low $A_9$ levels considered here, these
 `Sugawara-like' constraints $\mathfrak{L}$ transform as a linear
 representation of the upper parabolic subgroup  $\E^{+}$ generated
 by $\mathfrak{gl}(10)$ and the positive-root (raising) generators
 of $\E$. In addition, the latter representation can be embedded, at
 least at the levels considered here, and within the restriction to
 $\E^+$, into the integrable highest weight representation $L(\Lambda_1)$
 (to be defined below).
\end{itemize}

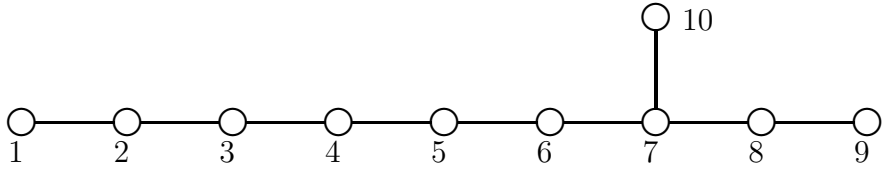
\begin{figure}
\begin{center}
\scalebox{1}{
\begin{picture}(340,60)
\put(5,-5){$1$} \put(45,-5){$2$} \put(85,-5){$3$}
\put(125,-5){$4$} \put(165,-5){$5$} \put(205,-5){$6$}
\put(245,-5){$7$} \put(285,-5){$8$} \put(325,-5){$9$}
\put(260,45){$10$} \thicklines
\multiput(10,10)(40,0){9}{\circle{10}}
\multiput(15,10)(40,0){8}{\line(1,0){30}}
\put(250,50){\circle{10}} \put(250,15){\line(0,1){30}}
\end{picture}}
\caption{\label{e10dynk}\sl Dynkin diagram of $\E$ with numbering
of nodes.}
\end{center}
\end{figure}

A key question at this point concerns the significance and the proper
interpretation of the constraints in the context of the $\E$ $\s$-model.
Because the level decomposition of $\E$ w.r.t. any of its regular
subgroups gives rise to an exponentially growing spectrum of degrees
of freedom, and because this proliferation of states may exceed by far
what would be needed to account for the space-time degrees of freedom
of the various maximal supergravities, and possibly even M theory,
it appears that suitable constraints may
be necessary in order to reduce their number to what is appropriate for
M theory. Furthermore it seems clear that the emergence of space (and time)
along the lines proposed in \cite{Damour:2002cu} cannot possibly be
explained without a proper understanding of the underlying constraints
on the $\s$-model dynamics.

The tensor structure of the constraints [cf. eqn.~(\ref{constdef})
below] coincides at levels $\ell =3,4$ and 5 with the tensor structure
of the so-called $L(\Lambda_1)$ representation of $\E$, while at level
$\ell =6$ it contains only one of the two irreducible Young tableaux
contained in $L(\Lambda_1)$. Let us recall that $L(\Lambda_1)$
is an integrable highest weight representation of $\E$
with Dynkin labels $[1\,0\,0\,0\,0\,0\,0\,0\,0\,0]$, where the `1' occurs
for the over-extended, hyperbolic node of the $E_{10}$ Dynkin diagram in
Fig.~\ref{e10dynk}.\footnote{\label{funwt}By definition, the
  fundamental weights  $\Lambda_i$ are dual to
 the simple roots of $E_{10}$, i.e. $\langle\Lambda_i|\alpha_j\rangle
 = \d_{ij}$ \cite{Kac}.} We note that the analogous representation
for $E_{11}$ had already appeared in previous work \cite{West:2003fc}.
The possible of occurrence of $L(\Lambda_1)$ in the present context might
therefore be interpreted as evidence for a covariant formulation along
the lines suggested there. However, when properly `contravariantised'
(in a sense to be explained in section~4.2), the constraints transform
covariantly only under the upper parabolic subgroup  $\E^{+}$ leaving
invariant the triangular gauge chosen for the representation of the
coset manifold; in particular, the putative highest weight state of the
representation is {\em not} annihilated by the relevant raising
operators. This is somewhat contrary to what one would expect
on the basis of a covariant formulation, as explained in
section~\ref{trmsec}. However, the transformations we obtain are
fully consistent with a Sugawara-type interpretation of the constraints.

The link between the canonical constraints obtained from supergravity
on the one hand, and a kind of Sugawara-like construction based on
$\E$ on the other hand, is the most remarkable result of the present
paper. It is not clear whether this fact indicates the existence of
a `covariant' set of equations whose gauge-fixed version would give
rise to the `one-dimensional' $E_{10}/K(E_{10})$ $\sigma$-model of
\cite{Damour:2002cu} supplemented by constraints as described in the
present paper. What seems clear is that such a putative `covariant'
formulation is likely to be of a rather unconventional type: in a scheme
with emergent space-time, the realisation of gauge symmetries must
necessarily differ from the standard realisation of gauge symmetries
in space and time. This would imply, for instance, that general covariance
and other space-time based gauge symmetries might emerge only together
with space-time itself, and thus not be fundamental, but only {\em emergent}
properties of the
theory.\footnote{In this context we may note that in canonical quantum
  gravity full general covariance likewise need not necessarily
  exist prior to the emergence of a
  classical space-time. In fact, no canonical quantisation of gravity is
  known, in which the full constraint algebra is realised {\em off shell},
  see e.g. \cite{Thiemann:2001yy,Nicolai:2005mc}. See also \cite{DN}
  for a related discussion.}

The evidence for a Sugawara-like construction for $\E$ presented here
is also noteworthy on the purely mathematical side. While the
existence of the  Sugawara construction for affine Lie algebras has
been known for a very  long
time~\cite{Sugawara:1967rw,Bardakci:1970nb,Goddard:1986bp,Kac}, no
analog for indefinite Kac--Moody algebras has ever been found. Nevertheless,
our results strongly indicate that such a generalisation does exist,
although it will certainly exhibit some unexpected features [as already
evident from the intricate tensor structure of the pertinent expressions].
Additional evidence for this conjecture derives from the fact that the
Sugawara-like structure of the coset constraints  reduces to
the known one when truncated to the affine $E_9$ subalgebra of $\E$.
As is well known, in the latter case we also have expressions bilinear in
the affine currents $L_m \propto \sum_n j^a_{m-n} j^a_n$ with the current
generators $j^a_n$~\cite{Sugawara:1967rw,Bardakci:1970nb,Goddard:1986bp}.
In the affine truncation of $E_{10}$ to $E_9$ (corresponding to a dimensional
reduction of maximal supergravity to two space-time dimensions), most
of the constraints `disappear', except for the diffeomorphism constraint,
denoted ${\cal{C}}^{(3)}$, an $SL(9)$ singlet. This singlet will be
shown to be directly related to the $L_{-1}$ Sugawara generator, which
is just the translation generator $(-d/dz)$ w.r.t. the spectral
parameter in a current algebra realisation of $E_9$. Via the
linear system of two-dimensional (super-)gravity \cite{BM} and its
hidden Virasoro symmetries \cite{Julia:1996nu}, diffeomorphisms in the
spectral parameter can be directly related to diffeomorphisms in the spatial
coordinate.

In summary, we would thus like to raise the possibility that the
Sugawara-like constraints $\mathfrak{L}$ given in section~4.2
constitute the beginnings of a generalisation of the affine Sugawara
construction for the hyperbolic Kac--Moody algebra $\E$, indicating
the existence of a so far undiscovered new structure inside
$\E$ and its envelopping algebra
(and possibly other hyperbolic Kac--Moody algebras) and hinting at the
existence of a more concrete realisation of these algebras analogous
to the current algebra realisation of affine algebras. In addition,
this generalisation might accommodate 10-dimensional spatial diffeomorphisms
in a similar way as the ordinary Sugawara construction realises
diffeomorphisms on the circle $S^1$. The present work could thus open
new avenues both towards analysing the hyperbolic $E_{10}$ algebra and
towards understanding how space (and time) emerge out of the geodesic
$\s$-model of \cite{Damour:2002cu}.

This article has the following structure. After introducing the
necessary notation for the $E_{10}/K(E_{10})$ model in
section~\ref{convsec} we propose a set of bosonic constraints and
fermionic constraints in
section~\ref{conssec}. After demonstrating their weak conservation
along the geodesic motion we show in section~\ref{dictsec} that they
coincide with the constraint equations of supergravity if the usual
$\E$/supergravity dictionary is used.
In section \ref{sugasec} we demonstrate how the bosonic constraints
can be reformulated in a Sugawara-like form and that the reduction to
$E_9$ gives the usual Sugawara construction. In this context we also
discuss the transformation properties of the constraints and show
that a parabolic subgroup $\E^+$ of $\E$ preserves the constraints.
In the concluding section we
return to the discussion of the original $E_{10}$ symmetry and the
interpretation of our results, including the relation to the Sugawara
construction.

\end{section}

\begin{section}{\bmath$E_{10}$\ubmath{}  model}
\label{convsec}

In this section we review the formalism of the $E_{10}/K(E_{10})$
coset model and fix our notations and conventions. We restrict attention
here to the bosonic fields and treat the fermions in section~\ref{susysec}.

\begin{subsection}{Coset variables and transformation}

We use the conventions of \cite{Damour:2004zy,Damour:2006xu} for
the $E_{10}$ commutation relations and for the construction of the
dynamics. Therefore, the time-dependent $E_{10}/K(E_{10})$ coset
element $\cV(t)$ gives rise to Lie algebra elements\footnote{We use
  $\E$ (and $K(\E)$) to  designate both the algebra and the group.} $\cQ\in
K(E_{10})$ and $\cP\in E_{10}\ominus K(E_{10})$ via the decomposition
\be
\p_t \cV\,\cV^{-1} = \cQ + \cP.
\ee
In terms of the generators at low $\mf{sl}(10)=A_9$ levels the `coset velocity'
$\cP$ and the `orthogonal' $K(E_{10})$ gauge connection $\cQ$ (which does
not enter the coset Lagrangian, see (\ref{action}) below) can be
expanded as
\bes{pqexp}\label{pexp}
\cP &=& \frac12 \cpd{0}{ab}S^{ab}
  + \frac1{3!}\cpd{1}{a_1a_2a_3}S^{a_1a_2a_3}
  + \frac1{6!}\cpd{2}{a_1\ldots a_6}S^{a_1\ldots a_6}\nn\\
&&\quad\quad\quad  + \frac1{9!}\cpd{3}{a_0|a_1\ldots a_8}
  S^{a_0|a_1\dots a_8} + \ldots,\\\label{qexp}
\cQ &=& \frac12 \cqd{0}{ab}J^{ab}
  + \frac1{3!}\cqd{1}{a_1a_2a_3}J^{a_1a_2a_3}
  + \frac1{6!}\cqd{2}{a_1\ldots a_6}J^{a_1\ldots a_6}\nn\\
&&\quad\quad\quad  + \frac1{9!}\cqd{3}{a_0|a_1\ldots a_8}
  J^{a_0|a_1\dots a_8} + \ldots.
\ees
where the indices $a,b,\ldots=1,\ldots,10$ are to be regarded as (`flat')
$SO(10)$ vector indices and the bracketed superscripts indicate
the $A_9$ level. The symmetric and anti-symmetric combinations $S$ and
$J$ of the $E_{10}$ generators are respectively defined by (on levels
$\ell=0,1,2,3$)
\begin{subequations}\label{sjdef}
\begin{align}
S^{ab} &= K^a{}_b + K^b{}_a\,,&
   J^{ab}&= K^a{}_b - K^b{}_a\,,&\\
S^{a_1a_2a_3} &= E^{a_1a_2a_3} + F_{a_1a_2a_3}\,,&
   J^{a_1a_2a_3} &= E^{a_1a_2a_3} - F_{a_1a_2a_3}\,,&\\
S^{a_1\ldots a_6} &= E^{a_1\ldots a_6} + F_{a_1\ldots a_6}\,,&
   J^{a_1\ldots a_6} &= E^{a_1\ldots a_6} - F_{a_1\ldots a_6}\,,&\\
S^{a_0|a_1\ldots a_8} &= E^{a_0|a_1\ldots a_8} + F_{a_0|a_1\ldots a_8}\,,&
   J^{a_0|a_1\ldots a_8} &= E^{a_0|a_1\ldots a_8} - F_{a_0|a_1\ldots a_8}\,,&
\end{align}
\end{subequations}
The elements $J$ generate the maximal compact subgroup
$K(E_{10})\subset E_{10}$ while the elements $S$ span the coset
$E_{10}\ominus K(E_{10})$ (which is not a subalgebra); their commutation
relations are given in \cite{Damour:2004zy,Damour:2006xu}.\footnote{With
  an overall minus
  sign correction in the $[\ell=-3,\ell=3]$ commutator compared to
  \cite{Damour:2004zy}.}

The coset has the usual non-linear symmetry transformations $\cV(t)\ra
k(t)\cV(t) g^{-1}$ with $g\in E_{10}$ a global rotation and $k(t)$ a
local gauge transformation, field dependent once a gauge is
chosen. Under this transformation one has
\be\label{trmlaw}
\cP \ra k \cP k^{-1} \,,\quad \cQ \ra k\cQ k^{-1} + \p_t k\,k^{-1}\,.
\ee
Note that $\cP$ is a $\K$-covariant object (coset representation of $\K$),
while $\cQ$ has the typical inhomogeneous transformation law of a gauge
connection of $\K$.
On the components defined in (\ref{pqexp}) this implies for instance
the following transformations for infinitesimal $\d k = \frac1{3!}
\clt{1}{c_1c_2c_3}J^{c_1c_2c_3} \in K(E_{10})$
\bes{dp}
\d_{\clt{1}{}}\cpd{0}{ab} &=&
\frac19\cld{1}{c_1c_2c_3}\cpd{1}{c_1c_2c_3}\d_{ab}
     -\cld{1}{c_1c_2a}\cpd{1}{bc_1c_2},\\
\d_{\clt{1}{}}\cpd{1}{a_1a_2a_3} &=& 3 \cld{1}{ca_1a_2}\cpd{0}{a_3c}
  +\frac1{6} \cld{1}{c_1c_2c_3}\cpd{2}{c_1c_2c_3a_1a_2a_3},\\
\d_{\clt{1}{}}\cpd{2}{a_1\ldots a_6} &=&-
  20\cld{1}{a_1a_2a_3}\cpd{1}{a_4a_5a_6}
  +\frac16\cld{1}{c_1c_2c_3}\cpd{3}{c_1|c_2c_3a_1\ldots
    a_6},\\\label{varp3}
\d_{\clt{1}{}}\cpd{3}{a_0|a_1\ldots a_8} &=&
  -56\big(\cld{1}{a_0a_1a_2}\cpd{2}{a_3\ldots a_8}
  -\cld{1}{a_1a_2a_3}\cpd{2}{a_4\ldots a_8a_0}\big)+\ldots.
\ees
In these equations we have employed a notational convention which we will
make use of throughout the remainder of this article. Namely, the r.h.s.
of the tensor equation is implicitly assumed to be projected onto the same
symmetry structure as the l.h.s., that is, the requisite symmetrisations
and antisymmetrisations are understood without being written out. For
example, the first two lines read in explicit form
\bes{antiexample}
\d_{\clt{1}{}}\cpd{0}{ab} &\equiv&
\frac19\cld{1}{c_1c_2c_3}\cpd{1}{c_1c_2c_3}\d_{ab}
     -\cld{1}{c_1c_2(a}\cpd{1}{b)c_1c_2}\,,\\
\d_{\clt{1}{}}\cpd{1}{a_1a_2a_3} &\equiv& 3 \cld{1}{c[a_1a_2}\cpd{0}{a_3]c}
  +\frac1{6} \cld{1}{c_1c_2c_3}\cpd{2}{c_1c_2c_3a_1a_2a_3}\,.
\ees
For later use we define the $K(E_{10})$ covariant derivative
\be\label{cD}
\cD \equiv \cD_t = \p_t - \cQ\,,
\ee
with the connection term $\cQ$ acting in the appropriate representation,
e.g. via commutators on $\cP$. The level zero generators $J^{ab}$ form an
$\mf{so}(10)$ subalgebra of $\mf{sl}(10)\subset E_{10}$ and we will
often use the $SO(10)$ covariant derivative
\be
D\equiv D_t = \p_t - \frac12 \cqd{0}{ab}J^{ab},
\ee
acting on representations of $SO(10)$; for example, for an
$\mf{so}(10)$ vector $v_a$ the covariant derivative evaluates to
$\cDso v_a = \p_t v_a - \cqt{0}{ab}v_b$.

\end{subsection}

\begin{subsection}{Equations of motion}

The equations of motion of the one-dimensional $E_{10}/K(E_{10})$
$\s$-model follow from the Lagrangian \cite{Damour:2002cu}
\be\label{action}
\cL = \frac1{4n} \langle \cP | \cP \rangle
\ee
with the lapse $n$ to ensure invariance under reparametrisations of $t$.
They are given by the geodesic equations
\be\label{eombos}
\cD_t \cP = \p_t \cP - \left[\cQ,\cP\right] = 0\,,
\ee
and the Hamiltonian constraint
\be\label{H}
{\cal H} (\cP) \equiv \langle\cP|\cP\rangle = 0
\ee
where for convenience we choose the gauge $n=1$ for the affine
parametrisation of the world-line. Imposition of (\ref{H}) requires
the coset space geodesic (\ref{eombos}) to be null.

A major simplification of (\ref{eombos}) is achieved by adopting the
(almost) {\em triangular gauge}, where $\cV$ depends only on the
level $\ell\geq 0$ degrees of freedom
\be\label{Vtriang}
\cV(t) = \cV_0(t) \exp\left[\frac1{3!} A_{mnp}(t) E^{mnp} +
        \frac1{6!} A_{m_1\dots m_6}(t) E^{m_1\dots m_6} + \dots \right]
\ee
As for instance explained in \cite{Damour:2002cu,Damour:2004zy}, the
first factor on the r.h.s. belongs to the $GL(10)$ subgroup of $\E$,
thus involving only the $\mathfrak{gl}(10)$ generators\footnote{The
 `dictionary' of section~3.3 associates this coset vielbein to the spatial
 zehnbein $e_m{}^a(t,{\bf x}_0)$ of $D=11$ supergravity evaluated at a fixed
 spatial point ${\bf x}_0$. Strictly speaking, we should notationally
 distinguish between the coset zehnbein (considered as a 10-by-10
 `submatrix' of $\cV(t)$), and the spatial zehnbein of supergravity,
 but we will refrain from doing so in order not to clutter up the
 notation.}
\be\label{v0}
\cV_0 (t) \equiv \exp \big(h_m{}^n (t) K^m{}_n\big)\,,\quad e_a{}^m =
(e^h)_a{}^m \,.
\ee
One should be careful here not to assign any special transformation
properties to $h_m{}^n$ appearing inside the exponential defining
$\cV_0$, whereas the exponentiated expression, that is $\cV_0$ itself,
does transform as a zehnbein, i.e. $\cV_0 \rightarrow k_0 \cV_0 g_0^{-1}$
with $k_0\in SO(10)$ and $g_0\in GL(10)$. By contrast, the higher level
fields appearing in the exponential inside (\ref{Vtriang}) do transform
as genuine $GL(10)$ tensors after the factor $\cV_0$ has been split off.
In other words, the indices $m,n,\dots$ in the second exponent of
(\ref{Vtriang}) can be thought of as `world' ($GL(10)$) indices in contrast to the
`flat' ($SO(10)$) indices $a,b,\dots$ in (\ref{pqexp}).
The fields at level one and two, respectively, correspond
to the 3-form field of $D=11$ supergravity, and its magnetic 6-form dual.
In this gauge, the `matrix'
$\cV(t)$ belongs to a parabolic subgroup of $\E$ which we designate by
$\E^+$. Furthermore, in this gauge, the connection
coefficients $Q^{(\ell)}$ appearing in (\ref{qexp}) are identified
(for $\ell\geq 1$) with the coset coefficients $P^{(\ell)}$ of
(\ref{pexp}) \cite{Damour:2004zy}
\be\label{triang}
\cqd{\ell}{} = \cpd{\ell} \qquad \mbox{for all $\,\ell\geq 1$}\,.
\ee
The commutators of \cite{Damour:2004zy} and the triangular gauge for
$\cV$ allow us to work out the following expressions for $\cP$ and $\cQ$
(cf. also \cite{Damour:2002cu})
\bes{panda}
\cqd{0}{ab} &=& e_{m[b}\p_t e_{a]}{}^m\quad,\quad
\cpd{0}{ab}=e_{m(b}\p_t e_{a)}{}^m\,,\\
\cpd{1}{a_1a_2a_3} &=& \frac12 e_{a_1}{}^{m_1}e_{a_2}{}^{m_2}
   e_{a_3}{}^{m_3}\p_t  A_{m_1m_2m_3}\, , \quad \textrm{etc.}
\ees
Here, the matrix $e_m{}^a$ is the inverse of $e_a{}^m$, {\em viz.}
$e_m{}^a e_a{}^n = \delta_m^n$.

The level decomposition allows us to decompose (\ref{eombos}) into
an infinite set of equations, which furthermore can be truncated
consistently by setting
\be\label{trunc3}
\cpd{\ell}{}=0\qquad \text{for}\; \ell>3\,.
\ee
With these gauge choices, and the truncation (\ref{trunc3}),
the Hamiltonian and the equations of motion (\ref{eombos}),
respectively, reduce to
\be
\langle\cP|\cP\rangle &=& \cpd{0}{ab}\cpd{0}{ab} - \cpd{0}{aa}\cpd{0}{bb}
 + \frac13 \cpd{1}{abc}\cpd{1}{abc} + \nn\\
&& + \frac1{4 \cdot 5!} \cpd{2}{a_1\dots a_6} \cpd{2}{a_1\dots a_6}
 + \frac2{9!} \cpd{3}{a_0|a_1\dots a_8} \cpd{3}{a_0|a_1\dots a_8} = 0
\ee
and
\bes{eomsa9}
\cDso \cpd{0}{ab} &=&
   -\frac19\d_{ab}\cpd{1}{c_1c_2c_3}\cpd{1}{c_1c_2c_3}
   + \cpd{1}{ac_1c_2}\cpd{1}{bc_1c_2}\nn\\
&& -\frac4{3\cdot6!}\d_{ab}\cpd{2}{c_1\ldots c_6}\cpd{2}{c_1\ldots c_6}
   +\frac{2}{5!}\cpd{2}{ac_1\ldots c_5}\cpd{2}{bc_1\ldots c_5}\nn\\
&& -\frac{2}{9!}\d_{ab}\cpd{3}{c_0|c_1\ldots c_8}\cpd{3}{c_0|c_1\ldots
     c_8}
  +\frac{16}{9!}\cpd{3}{c_0|ac_1\ldots c_7}\cpd{3}{c_0|bc_1\ldots
     c_7} \nn\\
&&   +\frac{2}{9!}\cpd{3}{a|c_1\ldots c_8}\cpd{3}{b|c_1\ldots c_8},\\
\label{eoml1}
\cDso\cpd{1}{a_1a_2a_3} &=&
   -3 \cpd{1}{ca_1a_2}\cpd{0}{ca_3}
   -\frac13 \cpd{1}{c_1c_2c_3}\cpd{2}{c_1c_2c_3a_1a_2a_3}\nn\\
&&   -\frac4{6!}\cpd{2}{c_1\ldots c_6}\cpd{3}{c_1|c_2\ldots
     c_6a_1a_2a_3}\\
\cDso\cpd{2}{a_1\ldots a_6} &=&
    6\cpd{2}{ca_1\ldots a_5}\cpd{0}{ca_6}
   -\frac13 \cpd{3}{c_1|c_2c_3a_1\ldots a_6}\cpd{1}{c_1c_2c_3}\\
\label{eoml3}
\cDso \cpd{3}{a_0|a_1\ldots a_8} &=&
  -\cpd{3}{c|a_1\ldots a_8}\cpd{0}{ca_0}
   + 8 \cpd{3}{a_0|ca_1\ldots a_7}\cpd{0}{ca_8}
\ees
Here, it is again understood that the r.h.s. is symmetrised in
accordance with the symmetries  on the l.h.s. of these equations,
as explained after (\ref{dp}).
For the level $\ell=3$ term in (\ref{eoml3}) this implies that
the r.h.s. vanishes if antisymmetrised over all nine free indices,
as required by the Young symmetries of the level three generator.
For clarity, we write out the contributions involving $\cpt{0}{}$
explicitly on the r.h.s., unlike in \cite{Damour:2004zy} where these
terms were absorbed into the derivative operator on the l.h.s. As we
will see below, however, the constraint analysis is simplified considerably
by re-absorbing these contributions into the derivative of a suitably
redefined l.h.s.

\end{subsection}

\end{section}

\begin{section}{Constraints}
\label{conssec}

We next show that the bosonic dynamical equation $\cD_t\cP=0$ (truncated at
levels $\ell \leq 3$) can be supplemented
by certain constraints ${\cal C}$ quadratic in the $\cP$, such that the
equations ${\cal C} \approx 0$ are all compatible with the dynamics of
the $E_{10}/K(E_{10})$ $\s$-model. Moreover, these constraints are
in one-to-one correspondence with the canonical constraints of supergravity,
as we shall see in the next section. Compatibility of constraints with
the equations of motions requires that the time derivatives of the
constraints vanish weakly (i.e. modulo the constraints) so that the
motion preserves the constraint surface determined by ${\cal C}(\cP)=0$.
In contrast to the Hamiltonian constraint (\ref{H}) which is an $\E$ singlet,
the constraints ${\cal C}(\cP)$ possess a more intricate structure with
regard to $\E$, which we shall now study.

\begin{subsection}{Bosonic constraints and weak conservation}
\label{bosconssec}

Motivated by the knowledge of the structure of the supergravity constraints
and of their `translation' in coset variables \cite{DHHS,Damour:2006xu},
we wish to study, purely from the viewpoint of the coset dynamics, the
possibility of imposing coset constraints $\cC \approx 0$
for a `constraint multiplet' of the general form
\bes{constdef}
\cCd{3}{a_1\ldots a_9} &=& \cpd{0}{ca_1}\cpd{3}{c|a_2\ldots a_9}
  + \alpha \cpd{1}{a_1a_2a_3}\cpd{2}{a_4\ldots a_9},\\
\cCd{4}{b_1\ldots b_{10}||a_1a_2} &=& \cpd{1}{a_1b_1b_2}\cpd{3}{a_2|b_3\ldots
    b_{10}}
  +\beta \cpd{2}{a_1b_1\ldots b_5}\cpd{2}{a_2b_6\ldots b_{10}},\\
\cCd{5}{b_1\ldots b_{10}||a_1\ldots a_5} &=& \cpd{2}{a_1\ldots a_4b_1b_2}
  \cpd{3}{a_5|b_3\ldots  b_{10}},\\
\cCd{6}{b_1\ldots b_{10}||a_0|a_1\ldots a_7} &=& \cpd{3}{a_0|b_1\ldots
  b_8} \cpd{3}{b_9|b_{10}a_1\ldots a_7}.
\ees
Let us clarify once more what various antisymmetrisations which are
understood here: for instance, all expressions are antisymmetric
in the 10-tuple of indices\footnote{The double lines $||$ in the
  subscripts of the constraints $\cCt{4}{}$, $\cCt{5}{}$ and
  $\cCt{6}{}$ serve as a mnemonic to separate the
  10-tuples $[b_1\dots b_{10}]$ from the other $SO(10)$ indices.}
 $[b_1\ldots b_{10}]$, as well as in the
indices $a_1 , a_2, \dots$, whereas the index $a_0$ is to be treated
separately (of course, the blocks of ten antisymmetric $b$ indices
could be eliminated by means of an $\eps$-symbol, but leaving them
explicit makes some of the structure more transparent). Thus, to give
one more example, the first equation in (\ref{constdef}) should be
read as follows,
\be
\cpd{0}{ca_1}\cpd{3}{c|a_2\ldots a_9} + \alpha
  \cpd{1}{a_1a_2a_3}\cpd{2}{a_4\ldots a_9}
\equiv \cpd{3}{c|[a_2\ldots a_9} \cpd{0}{a_1]c}
  + \alpha \cpd{1}{[a_1a_2a_3}\cpd{2}{a_4\ldots a_9]}\,.
\ee
The net effect of this prescription is that the l.h.s. and the r.h.s.
of all equations have the same symmetries, as it should be. Note
also that although $\cCt{6}{}$ could {\it a priori} contain {\em two}
irreducible Young tableaux (of $SL(10)$) in a specific linear combination,
the definition of $\cCt{6}{}$, together with the $\ell=3$ irreducibility
condition $P^{(3) [a_0|a_1...a_8]} = 0$ implies the algebraic constraint
${\cal C}^{(6)}_{b_1\dots b_{10}||[a_0|a_1\dots a_7]} = 0$. The ansatz
(\ref{constdef}) is motivated by previous studies of the supersymmetry
constraint in \cite{Damour:2006xu}.

As already mentioned in the introduction the tensor structure of the
flat indices appearing in (\ref{constdef}) is identical~\footnote{Except
  for the algebraic restriction on ${\cal C}^{(6)}$ just mentioned.}
with the one of
the lowest $SL(10)$ levels appearing in the $L(\Lambda_1)$
representation of $E_{10}$, with $3 \ell$ indices at each $A_9$ level
$\ell$ \cite{Kleinschmidt:2003jf}. In the form given in (\ref{constdef})
this is not entirely obvious: we must `remove' an $\epsilon$-symbol
with ten antisymmetric indices, counting the `missing' index in
${\cal C}^{(3)}$ as an extra (really: upper) index. In this way,
the index structure of the constraints becomes
\be
\cCd{3}{a} \, ,\; \cCd{4}{a_1a_2} \, , \;
\cCd{5}{a_1\dots a_5} \,,\; \cCd{6}{a_0|a_1\dots a_7} \, , \dots
\ee
which corresponds to the well known `central charge representations'
of maximal supergravity. The above pattern illustrates that, at least
for the low level representations displayed above, the relevant
$SL(10)$ Young tableaux are obtained, up to appropriate $\epsilon$
tensors, from the low level Young
tableaux of the adjoint of $E_{10}$ by removing one box in all
possible ways; so, for instance, the 3-form $[a_1a_2a_3]$ at level
one becomes a 2-form $[a_1a_2]$, and so on. However, at higher
levels there will appear extra representations. Similar representations
in the context of very-extended algebras have been studied
in \cite{West:2003fc,Kleinschmidt:2003jf}. The reason for introducing
the surplus antisymmetric indices in (\ref{constdef}) will be explained
in section~4.2, cf. remarks after (\ref{constdefnew}).

We find that demanding weak conservation  of the constraints above along the
coset motion, i.e. using the equations of motion (\ref{eomsa9}),
 {\em uniquely fixes} the numerical values of the
 coefficients $\alpha$, $\beta$ in (\ref{constdef}) to be:
 \be\label{abvals}
\alpha = 28 \,,\quad\beta=\frac{21}5.
\ee
With these special values, the result for
the time derivative of the constraints  is, using the
$SO(10)$-covariant derivative $D$ and setting $n=1$,
\bes{timederconst}
\cDso \cCd{3}{a_1\ldots a_9} &=& -9 \cpd{0}{ca_1}\cCd{3}{ca_2\ldots a_9}
  + 10 \cpd{1}{c_1c_2c_3}\cCd{4}{a_1\ldots a_9 c_1||c_2c_3}\\
&&  -\frac7{36}\cpd{2}{c_1\ldots c_6}\cCd{5}{a_1\ldots a_9c_1||c_2\ldots
    c_6}
  + \frac{160}{9!} \cpd{3}{c_0|c_1\ldots c_8}\cCd{6}{a_1\ldots
    a_9c_1||c_0|c_2\ldots c_8},\quad\quad\mbox{ }\nn\\
\cDso \cCd{4}{b_1\ldots b_{10}||a_1a_2} &=& -10
\cpd{0}{cb_1}\cCd{4}{cb_2\ldots b_{10}||a_1a_2}
 -2  \cpd{0}{ca_1}\cCd{4}{b_1\ldots b_{10}||ca_2} \\
&& -\frac56 \cpd{1}{c_1c_2c_3}\cCd{5}{b_1\ldots b_{10}||c_1c_2c_3a_1a_2}
+\frac3{5!}\cpd{2}{c_1\ldots c_6}
  \cCd{6}{b_1\ldots b_{10}||c_1|c_2\ldots c_6 a_1a_2},\nn\\
\label{timederconst5}
\cDso \cCd{5}{b_1\ldots b_{10}||a_1\ldots a_5} &=&
  -10 \cpd{0}{cb_1}\cCd{5}{cb_2\ldots b_{10}||a_1\ldots a_5}
  -5  \cpd{0}{ca_1}\cCd{5}{b_1\ldots b_{10}||ca_2\ldots a_5}\nn\\
&&  -\frac2{15} \cpd{1}{c_1c_2c_3}\cCd{6}{b_1\ldots b_{10}||c_1|c_2c_3a_1\ldots
    a_5},\\
\cDso \cCd{6}{b_1\ldots b_{10}||a_0|a_1\ldots a_7} &=&
  -10\cpd{0}{cb_1} \cCd{6}{cb_2\ldots b_{10}||a_0|a_1\ldots a_7}
  - \cpd{0}{ca_0} \cCd{6}{b_1\ldots b_{10}||c|a_1\ldots a_7}\nn\\
&& -7\cpd{0}{ca_1}\cCd{6}{b_1\ldots b_{10}||a_0|ca_2\ldots a_7}.
\ees
again with all required symmetrisations implied.
Because the time derivatives of the constraints are again proportional to
the constraints, the constraints are {\em weakly conserved} in this
truncation, hence the constraints can be imposed to yield a consistent
restriction of the dynamics as claimed.

These weak conservation equations exhibit two remarkable structures:
$(i)$ the universal appearance of the negative of the zero-level coset
velocity $- \cpt{0}{ab}$ acting (by being contracted) on the r.h.s., on
each index of $\cCt{\ell}{\ldots b \ldots}$, and $(ii)$ a {\em triangular
structure} of the terms on the r.h.s. involving the  $\cCt{\ell'}{}$'s
with $\ell'$ differing from the level $\ell$ appearing on the l.h.s..
This triangular structure is reminiscent of a highest (or lowest) weight
representation in that the time derivatives
$D\cCt{\ell}{}$ involve only constraints $\cCt{\ell'}{}$ with
levels $\ell'\geq \ell$,  multiplied by $\cpt{\ell'-\ell}{}$.

We shall show below how these two remarkable structural elements
of the above weak-conservation equations are connected to a
Sugawara-like reformulation of the constraints. For the time being,
we only note that the conservation of the constraints implies
that one can consistently constrain null geodesic motion on
$E_{10}/K(E_{10})$ beyond the null geodesic constraint, at least in
the truncation (\ref{trunc3}) and in triangular gauge. Note also
that the Hamiltonian constraint is not required for the
above conservation equations to hold.

\end{subsection}

\begin{subsection}{The supersymmetry constraint}
\label{susysec}

The results of the preceding sections can be generalised to the case
where spin degrees of freedom are added, supplementing the bosonic
constraints by a supersymmetry constraint corresponding to local
supersymmetry. The inclusion of the fermionic fields of supergravity
has already been studied from an $E_{10}$ point of view in
\cite{Damour:2005zs,de Buyl:2005mt,de Buyl:2005zy,Damour:2006xu}.

$K(E_{10})$ possesses an unfaithful spinor representation $\psi_a$ of
dimension $320$ which transforms as a vector-spinor under
$SO(10)\subset K(E_{10})$ \cite{Damour:2005zs,de Buyl:2005mt}. The
$K(E_{10})$ covariant equation of motion for this representation is
the $K(E_{10})$ Dirac equation
\be\label{eomvs}
\cD_t \psi_a &=& \cDso_t \psi_a
  -\frac1{12}\cqd{1}{b_1b_2b_3}\G^{b_1b_2b_3}\psi_a
  -\frac23\cqd{1}{ab_1b_2}\G^{b_1}\psi^{b_2}
  +\frac16\cqd{1}{b_1b_2b_3}\G_a{}^{b_1b_2}\psi^{b_3}\nn\\
&& -\frac1{2\cdot 6!}\cqd{2}{b_1\ldots b_6}\G^{b_1\ldots
    b_6}\psi_a
   -\frac1{180}\cqd{2}{b_1\ldots b_6}  \G_a{}^{b_1\ldots
     b_5}\psi^{b_6}
   +\frac1{72}\cqd{2}{ab_1\ldots  b_5}\G^{b_1\ldots
     b_4}\psi^{b_5}\nn\\
&& -\frac2{3\cdot 8!}\left( \cqd{3}{b_0|b_1\ldots b_8}\G_a{}^{b_1\ldots
    b_8}\psi^{b_0} +8 \cqd{3}{a|b_1\ldots b_8}\G^{b_1\ldots
    b_7}\psi^{b_8} \right.\nn\\
&& \quad\quad\quad\quad\left.
   +2\cqd{3}{c|cb_1\ldots b_7}\G^{b_1\ldots b_7}\psi_a
  -28\cqd{3}{c|cb_1\ldots b_7}\G_a{}^{b_1\ldots
    b_6}\psi^{b_7}\right)\,.
\ee
The $\G$-matrices are real $(32\times 32)$-matrices of $SO(10)$.
In the triangular gauge (\ref{triang}), which we use throughout, we
can replace the connection coefficients $Q^{(\ell)}$ appearing in this
$K(E_{10})$ covariant derivative by the coset coefficients $P^{(\ell)}$.
Using the dictionary (\ref{a9dictionary}) one can then verify that --
modulo higher order gradients, as always -- the above equation
coincides with the Rarita Schwinger equation of maximal supergravity
\cite{Damour:2005zs,de Buyl:2005mt}.

Because one can supplement the bosonic coset dynamics (\ref{eomsa9})
by the weakly conserved constraints (\ref{constdef}) it is natural
to ask if similarly a supersymmetry constraint exists in the
combined bosonic and fermionic system which is weakly conserved. A
candidate constraint was described in \cite{Damour:2006xu} where it
was derived from supergravity. It has the form $\cS\approx 0$, with
\be
\label{susycons}
\G_0\cS &=&
  \frac12\left(\cpd{0}{ab}\G^a-\cpd{0}{cc}\G_b\right)\psi^b
  + \frac14 \cpd{1}{c_1c_2c_3}\G^{c_1c_2}\psi^{c_3}
  -\frac1{2\cdot 5!}\cpd{2}{c_1\ldots c_6}\G^{c_1\ldots
   c_5}\psi^{c_6}\nn\\
&& +\frac1{6\cdot 6!}\cpd{3}{b|bc_1\ldots c_7}\G^{c_1\ldots
  c_6}\psi^{c_7}
-\frac1{3\cdot 8!}\cpd{3}{b|c_1\ldots c_8}
  \G^{c_1\ldots c_8}\psi^b\,.
\ee
This expression
was shown in \cite{Damour:2006xu} to coincide with the appropriately
truncated supersymmetry constraint of supergravity upon use of
(\ref{a9dictionary}) (and, in fact, can be used to re-derive the
dictionary).

We can work out the time derivative of (\ref{susycons}) using solely
the $K(E_{10})$ covariant Dirac equation (\ref{eomvs}) to
obtain
\be\label{DS}
D_t \cS &=& \left[ \frac1{12} \cpd{1}{a_1a_2a_3} \G^{a_1a_2a_3} +
 \frac1{2\cdot 6!} \cpd{2}{a_1\dots a_6} \G^{a_1\dots a_6}
 +\frac4{3\cdot 8!} \cpd{3}{b|ba_1\ldots a_7}\G^{a_1\ldots a_7}
  \right]\cdot \cS      \nn\\
&& + \frac{3}{8!} \cCd{3}{a_1 \dots a_9}\G^0 \G^{a_1 \dots a_8} \psi^{a_9}
   -\frac1{8!} \cCd{4}{b_1\ldots b_{10}||a_1a_2}
      \G^0 \G^{b_1\ldots b_{10}}\G^{a_1}\psi^{a_2}\nn\\
&&   -\frac5{3\cdot 8!} \cCd{4}{b_1\ldots b_{10}||a_1a_2}
       \G^0 \G^{b_1\ldots b_9}\G^{a_1a_2}\psi^{b_{10}}
  +\frac{5}{8\cdot 9!} \cCd{5}{b_1\ldots b_{10}||a_1\ldots
    a_5}\G^0\G^{b_1\ldots b_9}\G^{a_1\ldots a_5}\psi^{b_{10}}\nn\\
&& - \frac{15}{16\cdot 9!} \cCd{5}{b_1\ldots b_{10}||a_1\ldots
    a_5}\G^0\G^{b_1\ldots b_{10}}\G^{a_1\ldots a_4}\psi^{a_5}+\ldots
\ee
Consequently, the supersymmetry constraint is also conserved on the constraint
surface where one imposes both the bosonic constraints (\ref{constdef})
and the supersymmetry constraint (\ref{susycons}) itself. This result
does not depend on the Hamiltonian constraint, and thus
provides no extra consistency checks on the latter.

Remarkably, the above conservation equation can be recast in terms
of a $K(E_{10})$ covariant derivative acting on $\cS$, suggesting
that the supersymmetry constraint behaves as a $K(E_{10})$ Dirac spinor.
This is achieved by shifting the first three terms on the r.h.s. of
(\ref{DS}) to the l.h.s. In triangular gauge we can then replace
$P^{(\ell)}$ by $Q^{(\ell)}$ to obtain
\be\label{DS1}
\cD_t \cS &\equiv& \left[ D_t - \frac1{12} \cqd{1}{a_1a_2a_3} \G^{a_1a_2a_3} -
 \frac1{2\cdot 6!} \cqd{2}{a_1\dots a_6} \G^{a_1\dots a_6}
 - \frac4{3\cdot 8!} \cqd{3}{b|ba_1\ldots a_7}\G^{a_1\ldots a_7}
  + \dots
\right] \cS  \nn\\
&= &   \frac{3}{8!} \cCd{3}{a_1 \dots a_9}\G^0 \G^{a_1 \dots a_8} \psi^{a_9}
   -\frac1{8!} \cCd{4}{b_1\ldots b_{10}||a_1a_2}
       \G^0\G^{b_1\ldots b_{10}}\G^{a_1}\psi^{a_2}\nn\\
&&   -\frac5{3\cdot 8!} \cCd{4}{b_1\ldots b_{10}||a_1a_2}
        \G^0\G^{b_1\ldots b_9}\G^{a_1a_2}\psi^{b_{10}}
  +\frac{5}{8\cdot 9!} \cCd{5}{b_1\ldots b_{10}||a_1\ldots
    a_5}\G^0\G^{b_1\ldots b_9}\G^{a_1\ldots a_5}\psi^{b_{10}}\nn\\
&& - \frac{15}{16\cdot 9!} \cCd{5}{b_1\ldots b_{10}||a_1\ldots
    a_5}\G^0\G^{b_1\ldots b_{10}}\G^{a_1\ldots a_4}\psi^{a_5}+\ldots\,,
\ee
where the dots indicate possible higher level contributions.
In this respect the supersymmetry constraint thus behaves like a
Dirac spinor representation of $K(E_{10})$ (defined in
\cite{de Buyl:2005zy,Damour:2005zs,Damour:2006xu}).
However, as already noted in \cite{Damour:2006xu} the
supersymmetry constraint  does not transform properly under $K(E_{10})$
(this observation is analogous to the one which will be made in
section~\ref{trmsec}
for the bosonic constraints), so the significance of the appearance of
the $K(E_{10})$ covariant derivative in the above equation remains
to be clarified.

\end{subsection}

\begin{subsection}{Translation to supergravity}
\label{dictsec}

In the previous section we have worked entirely within the coset model,
except for the fact that we {\em motivated} the general algebraic
structure (without using precise information about the numerical
coefficients $\a$ and $\beta$ in (\ref{constdef})) of possible
constraints by previous knowledge from the supergravity side of the
coset/supergravity correspondence. In this section we shall use the
`dictionary' relating the unconstrained $\s$-model to the dynamical
equations of supergravity \cite{Damour:2002cu} to compare the constraints
(\ref{constdef}) to the known canonical constraints of supergravity
in detail. It is therefore  gratifying that we shall re-obtain the uniquely
determined coset values (\ref{abvals}) by matching the supergravity
expressions to the coset ones in this way.

All bosonic equations are displayed in table~\ref{eqtab}. Both the
dynamical equations and the constraint equations can be obtained by
the standard ADM procedure from the field equations of $D=11$
supergravity \cite{CJS}, namely the Einstein equations $\cG_{AB}$ and
the matter (4-form field strength) equations $\cM^{BCD}$; in the
present conventions the latter read
\bes{d11eom}
\cG_{AB} &=& R_{AB}-\frac13 F_{ACDE} {F_B}^{CDE}
           + \frac1{36} \eta_{AB} F_{CDEF} F^{CDEF}\,,\\\label{feq}
\cM^{BCD} &=& D_A F^{ABCD}+\frac1{576} \eps^{BCDE_1\dots E_4 F_1\dots F_4}
                  F_{E_1\dots E_4} F_{F_1\dots F_4}\,,
\ees
where indices are flat space-time indices $A,B=0,1,\ldots,10$.

\begin{table}[t]
\centering
\begin{tabular}{|c|c|l|}
\hline
Supergravity & Coset & Name\\
\hline\hline
$\cG_{ab} =0$ & $\cD \cpd{0}{ab} = 0$ & Einstein dynamical eq.\\
$\cM^{a_1a_2a_3} =0$ & $\cD \cpd{1}{a_1a_2a_3} = 0$ &Matter dynamical eq.\\
$D_{[0} F_{a_1\ldots a_4]} = 0$ & $\eps_{a_1\ldots a_4b_1\ldots
    b_6}\cD\cpd{2}{b_1\ldots b_6} =0$ & F-Bianchi I\\
$R_{[0a\,b]c} =0$ & $\eps_{bcd_1\ldots d_8} \cD \cpd{3}{a|d_1\ldots d_8} =0$
  & R-Bianchi I\\
\hline
$\cG_{00} = 0$ & $\langle \cP|\cP\rangle =0$ & Hamiltonian constraint\\
$\cG_{0a} =0$ & $\eps_{ac_1\ldots c_9}\cCd{3}{c_1\ldots c_9} =0$ &
  Momentum constraint\\
$\cM^{0a_1a_2} =0$ & $\eps_{b_1\ldots b_{10}}\cCd{4}{b_1\ldots
  b_{10}||a_1a_2} = 0$ & Gauss constraint\\
$D_{[c_1}F_{c_2\ldots c_5]} =0$ & $\eps_{b_1\ldots b_{10}}\eps_{a_1\ldots
  a_5c_1\ldots c_5} \cCd{5}{b_1\ldots b_{10}||a_1\ldots a_5}=0$ &
  F-Bianchi II\\
$R_{[c_1c_2\,c_3]a_0} =0$ & $\eps_{b_1\ldots b_{10}}\eps_{a_1\ldots
  a_7c_1c_2c_3} \cCd{6}{b_1\ldots b_{10}||a_0|a_1\ldots a_7} =0$ &
  R-Bianchi II\\
\hline
\end{tabular}
\caption{\label{eqtab}\em Complete list of all (bosonic) coset equations
  and their corresponding (bosonic) supergravity equations.}
\end{table}

In order to make the comparison of the supergravity equations to the
coset equations we need to gauge-fix and truncate the supergravity
model.
More specifically, using the
conventions of ref.~\cite{Damour:2006xu}, the `dictionary' is specified
by assuming a zero-shift gauge of the vielbein, {\em viz.}
\be
E_M{}^A = \left(\begin{array}{cc}N&0\\0&e_m{}^a\end{array}\right) \;\; ;
\ee
we will write $e\equiv \det(e_m{}^a)$ for the determinant of the spatial
zehnbein. All indices $a,b,\dots$ here and in (\ref{a9dictionary})
below are flat $SO(10)$ indices. Furthermore, all supergravity fields
are evaluated at a fixed spatial
point ${\bf x}_0$, and are truncated by setting spatial frame derivatives
of the spin connection, the field strengths and the lapse to zero:
$\p_a\omega_{b\,cd}=\p_a F_{0bcd} =\p_a F_{bcde} = \p_a N\equiv 0$.
The coefficients of anholonomy $\O_{ab\,c}$ are assumed to be tracefree,
i.e. $\Omega_{ab\,b} = 0$. The dictionary is then given by
\bes{a9dictionary}
n(t) &\lrra& N e^{-1} (t,{\bf x}_0)\,,\\
\cqd{0}{ab}(t) &\lrra& -N \o_{0\,ab}(t,{\bf x}_0)
 \equiv - e_{[a}{}^m \partial_t e_{m|b]}(t,{\bf x}_0)     \,,\\
\cpd{0}{ab}(t) &\lrra& -N \o_{a\,b0}(t,{\bf x}_0)
  \equiv - e_{(a}{}^m \partial_t e_{m|b)}(t,{\bf x}_0) \,,\\
\cpd{1}{a_1a_2a_3}(t) &\lrra& N F_{0a_1a_2a_3}(t,{\bf x}_0)\,,\\
\cpd{2}{a_1\ldots a_6}(t) &\lrra& -\frac1{4!} N \eps_{a_1\ldots
  a_6b_1\ldots b_4}F_{b_1\ldots b_4}(t,{\bf x}_0)\,,\\
\cpd{3}{a_0|a_1\ldots a_8}(t) &\lrra& \frac34 N \eps_{a_1\ldots
  a_8b_1b_2}\tO_{b_1b_2\, a_0}(t,{\bf x}_0)\,,
\ees
where $a,b,\dots$ are now to be interpreted as {\em flat} spatial
indices in ten spatial dimensions.

Substituting the above expressions into the constraints (\ref{constdef})
with the values (\ref{abvals}), and contracting with an $\eps$-tensor
we arrive at
\bes{sugraconst}
N^{-2}\eps_{a_1\ldots a_9\,a} \cCd{3}{a_1\ldots a_9} &=&
  \frac12 \,8!\left(3\tO_{ab\,c}\o_{b\,c0} +
  F_{ab_1b_2b_3}F_{0b_1b_2b_3}\right),\\
N^{-2}\eps_{b_1\ldots b_{10}}\cCd{4}{b_1\ldots b_{10}||a_1a_2} &=&
  \frac32 \, 8!\bigg(F_{0a_1b_2b_2}\tO_{b_1b_2\,a_2} \\
&&\quad\quad\quad  + \frac1{576}\eps_{a_1a_2b_1\ldots b_4c_1\ldots
    c_4} F_{b_1\ldots   b_4} F_{c_1\ldots c_4}\bigg),\nn\\
N^{-2}\eps_{b_1\ldots b_{10}}\eps_{a_1\ldots a_5 c_1\ldots c_5}
  \cCd{5}{b_1\ldots b_{10}||a_1\ldots a_5} &=& -6!\cdot 8!\;
  \tO_{c_1c_2\,d}F_{dc_3c_4c_5},\\
N^{-2} \eps_{b_1\ldots b_{10}}\eps_{a_1\ldots a_7c_1c_2c_3}
  \cCd{6}{b_1\ldots b_{10}||a_0|a_1\ldots a_7} &=& 9\cdot 7!\cdot
  7!\cdot 3! \; \tO_{c_1c_2\,d}\tO_{c_3d\,a_0}.
\ees
These expressions correspond to the truncated versions of the
supergravity constraints with the truncation as specified above.
In the order given in (\ref{sugraconst}) the
supergravity constraint equations are: the momentum (diffeomorphism)
constraint, the Gauss constraint, the F-Bianchi constraint and the
R-Bianchi constraint (cyclic identity $R_{[c_1c_2\,c_3]a_0}=0$). The
truncation here amounts to ignoring spatial gradients of the spin
connection and field strength terms (for instance, the full momentum
constraint would have an extra term $\propto \partial_b \omega_{b\, a0}$,
and the Gauss constraint an extra $\partial_c F_{0abc}$). It is a
non-trivial fact that the same numerical values (\ref{abvals}) for
$\alpha$ and $\beta$ ensure  the weak
conservation of the constraints both w.r.t. the coset dynamics and the
supergravity one, because the two Hamiltonians do differ at level 3
(even within the truncation we use on both sides) by a term that could
have modified the weak conservation condition (but did not). [See
  \cite{Damour:2004zy} for the precise mismatches in $\cpt{3}{}$ terms.]

The equations
given in table~\ref{eqtab} exhaust all bosonic equations of the $D=11$
supergravity
system and we have found appropriate $\E$ counterparts in the present
truncation.\footnote{The Riemann
  Bianchi components
  $R_{[0a\,b]0}$ and $R_{[ab\,c]0}$ vanish identically in our
  truncation. In the full gravity theory the relation
  $R_{[abc]0}=-R_{[0a\,b]c}$ holds which seems to be inconsistent with
  R-Bianchi I. The resolution is that such a relation no longer holds
  in the truncation appropriate for the $\s$-model ($E_{10}$ does not
  know about the Riemann tensor).}

\end{subsection}

\end{section}

\begin{section}{A Sugawara-like construction for $\E$?}
\label{sugasec}

In this section we investigate in more detail the structure and
properties of the bosonic constraints (\ref{constdef}) and show
that they can be equivalently expressed in a Sugawara-like form
$\cJ\otimes\cJ$ in terms of the $\E$ Noether current $\cJ$. In the
level-3 truncation we will see that the constraints, when written
in this form, transform covariantly under a Borel subgroup $\E^+\subset\E$.
In the last subsection, we will establish the link with the more
familiar affine Sugawara construction by considering the embedding
$E_9\subset\E$.

\begin{subsection}{The $\E$ Noether Current}

By Noether's theorem, the $\E$ global symmetry  of the coset action
implies the existence of an infinite number of exactly conserved
quantities, {\em viz.}~\footnote{By abuse of language, we usually refer
to $\cJ$ as the `{\em (conserved) current}', although one should more
properly speak of a conserved charge.}
 \be\label{J}
\cJ = n^{-1}\cV^{-1}\cP\cV\,.
\ee
Henceforth (as elsewhere in this paper) we will use the gauge $n=1$.
The current $\cJ$  takes values in the Lie algebra of $\E$, and is
time-independent, that is, the $\s$-model equations of motion (\ref{eombos})
are equivalent to current conservation $\p_t\cJ=0$. Expanding the current
according to level and making use of the triangular gauge (\ref{Vtriang})
for $\cV(t)$, it is straightforward to see that the level truncation
condition (\ref{trunc3}) is equivalent to
\be\label{Jt}
\cJ^{(\ell)} = 0 \qquad \mbox{for $\ell = -4, -5,-6,\dots$}
\ee
Consequently, we have the expansion
\be\label{Jtrunc}
\cJ &=& \frac1{9!}\cjd{-3}{m_0|m_1\dots m_8} F_{m_0|m_1\dots m_8} +
        \frac1{6!}\cjd{-2}{m_1\dots m_6} F_{m_1\dots m_6} +
        \frac1{3!}\cjd{-1}{mnp} F_{mnp} + \nn\\
    && + \stackrel{(0)}{J}{}^n{}_m  K^m{}_n + \frac1{3!}\cjo{1}{mnp} E^{mnp}
       + \frac1{6!} \cjo{2}{m_1\dots m_6} E^{m_1\dots m_6} + \dots
\ee
where the ellipses on the right stand for infinitely many non-vanishing
positive-level components of $\cJ$. Expressing the current components
in terms of (contravariant) velocities and fields, we obtain at the
lowest levels
\bes{jandp}
\cjd{-3}{m_0|m_1\dots m_8}&=& P^{(3)m_0|m_1\dots m_8} \,,\\
\cjd{-2}{m_1\dots m_6} &=& P^{(2)m_1\dots m_6}
  + \frac1{3!} A_{pqr} P^{(3) p|qrm_1\dots m_6}   \,,\\
\cjd{-1}{mnp} &=& P^{(1) mnp} + \frac1{3!} A_{rst} P^{(2)rstmnp} + \nn\\
 && +  \left(\frac23 A_{r_1\dots r_6} + \frac1{72} A_{r_1r_2r_3} A_{r_4r_5r_6}
  \right) P^{(3) r_1|r_2 \dots r_6 mnp}   \,.
\ees

Three important features here should be noted:
\begin{itemize}
\item $\cJ$ is an $\E$ object, transforming as
 $\cJ \, \rightarrow \, \cJ' = g \cJ g^{-1}$ under rigid $\E$
 transformations $g\in\E$. This implies in particular that all indices
 in eq.~(\ref{Jtrunc}) are $GL(10)$ (`world') indices, which
 are covariant or contravariant according to their position, as
 indicated in the above formula.
\item $\cJ$ is manifestly inert under $K(\E)$. This means that the
 truncation condition (\ref{Jt})
 is {\em gauge invariant}, hence does {\em not} rely on any particular
 choice of gauge (such as the triangular gauge). In contrast, the
 truncation condition (\ref{trunc3}) on $\cP$ is not gauge invariant.
\item Unlike the velocities $P^{(\ell)}$, of which there are only four
 non-vanishing components (for $\ell=0,1,2,3$) with the truncation
 (\ref{trunc3}), there are {\em infinitely many} non-vanishing
 components $J^{(\ell)}$ at positive level. As shown in \cite{Damour:2002et}
 and explicitly exhibited in (\ref{jandp}) the most negative level
 component of $\cJ$ is purely {\it velocity}- (or {\it momentum}-)like,
 while the positive level components contain an increasing dependence
 on the coset coordinates (or {\it positions})
 $ = \{ A_{mnp}, A_{m_1\dots m_6}, A_{m_0|m_1\dots m_8} , \dots\}$.
 An Euclidean-group analog of this situation  would be to consider
 geodesic motion on Euclidean space: the conserved quantities would
 be $(p_i, L_{ij})$, where the linear momentum $p_i$ is pure velocity,
 whereas the angular momentum $L_{ij}$ involves both velocities and
 positions.
\end{itemize}

Clearly, the truncation condition (\ref{Jt}) is preserved only by
the parabolic subgroup $\E^+ \subset\E$ which is generated by the level
$\ell\geq 0$ generators of $\E$, that is, by ${\mathfrak{gl}}(10)$ and the
positive level generators $E^{mnp}, E^{m_1\dots m_6}, \ldots$. This is
the part  of $\E$ which transforms the coordinates, but leaves unchanged the
coset velocoities (or momenta). This property of $\E^+$ is one of the
reasons why the (presently known) coset constraints will transform only
under $\E^+$; indeed, any negative level transformations
will automatically violate (\ref{Jtrunc}). Under such a (strictly
upper triangular) transformation
\be
g= \exp \left( \frac1{3!} \Lambda^{(1)}_{mnp} E^{mnp} + \frac1{6!}
\Lambda^{(2)}_{m_1\dots m_6} E^{m_1\dots m_6} + ... \right)\in\E^+
\ee
the lowest components of the current transform as
\bes{Jtrafo}
\delta \cjd{0}{m}{}_n =&=&
  \frac1{18} \delta^m_n \Lambda^{(1)}_{pqr} \cjd{-1}{pqr} -
  \frac12 \Lambda^{(1)}_{npq} \cjd{-1}{mpq}   \,, \\
\delta \cjd{-1}{mnp} &=& \frac16 \Lambda^{(1)}_{qrs} \cjd{-2}{qrsmnp} \,,\\
\delta \cjd{-2}{m_1\dots m_6} &=&
    \frac16 \Lambda^{(1)}_{qrs} \cjd{-3}{q|rsm_1\dots m_6}  \,, \\
\delta \cjd{-3}{m_0|m_1\dots m_8}  &=& 0\,.
\ees
Here, we have shown the infinitesimal result when  only
$\Lambda^{(1)}_{pqr}$ is non-zero. The truncation (\ref{Jt}) implies
that $J^{(-3)}$ is invariant. Infinitesimally we have in general that
$\d J^{(\ell)}=\sum_n \Lambda^{(n)}J^{(\ell-n)}$.

We shall next investigate the relation of the conserved charges $\cJ$ to
the constraints derived in the foregoing section. Before doing so, however,
it is useful to recall that there is already one constraint which can be
expressed in manifest `Sugawara form', the Hamiltonian constraint.
Namely, from (\ref{J}) it is evident that
\be
\langle\cP | \cP\rangle = 0 \qquad \Longleftrightarrow \qquad
\langle\cJ | \cJ\rangle = 0 \, .
\ee
Furthermore this constraint is obviously an $\E$ singlet.

\end{subsection}

\begin{subsection}{Sugawara-like construction of the constraints}
\label{sugsec}

As mentioned at the end of section~\ref{bosconssec} above, the weak
conservation of the constraints exhibits two remarkable structural features.
The first of these is the universal action of the zero-level coset
velocity $-\cpt{0}{ab}$ on the r.h.s. of the weak conservation equations
(\ref{timederconst}). Namely, as already observed in \cite{Damour:2004zy},
this action  can be combined with the similar universal action of the
$SO(10)\subset\K$ gauge  connection $-\cqt{0}{ab}$ on the l.h.s. by
using the formulas (\ref{a9dictionary})~\footnote{We always adhere to
the conventions and notations of \cite{Damour:2006xu}.}
\be
\cpd{0}{ab} - \cqd{0}{ab} = - e_b{}^m \partial_t e_{ma}
   = + e_{ma} \partial_t e_b{}^m
\ee
whence
\be
\partial_t v_a + \big( \cpd{0}{ab} - \cqd{0}{ab} \big) v^b
  = e_{ma} \partial_t v^m \, .
\ee
Here, $v^m\equiv e_a{}^m v^a$ is the {\em contravariant} `world'
version of  the tangent space vector $v^a\equiv v_a$. The (inverse)
coset zehnbein $e_m{}^a=(e^{-h})_m{}^a$ is obtained from $\cV_0$ as in
(\ref{v0}).
Therefore, the universal structure of the $\cpt{0}{ab}$ and
$\cqt{0}{ab}$ contributions in the weak conservation equation
(\ref{timederconst}) above
is precisely such that they can all be eliminated if one replaces
all the $SO(10)$ `flat' indices $a,b, \ldots$ by {\em contravariant}
`world' $GL(10)$ indices $m,n,\ldots$
Accordingly, we can now convert the constraints of the previous section
(written in terms of `flat' indices) to {\em contravariant form} by
means of the coset zehnbein defining
\be\label{glconv}
\cCid{-3}{m_1\ldots m_9} \equiv e^{a_1m_1}\cdots e^{a_9m_9}
 \cCd{3}{a_1\ldots a_9}\;\; , \quad \text{etc.}
\ee
In this `contravariant'
form the constraints $\mathfrak{C}$ are now $GL(10)$ tensors rather
than $SO(10)$
(`flat' or `Lorentz') tensors. The reasons for switching to a labelling
with {\em negative} integers will become apparent shortly. The
conversion (\ref{glconv}) into $GL(10)$ world indices only changes the
transformation under the level $\ell=0$; below we will see that
converting all $\K$ indices into $\E$ indices is more natural and
gives a more unified structure.

The second noteworthy feature was the triangular evolution structure
of (\ref{timederconst}), which becomes {\em strictly upper triangular}
with the above redefinitions; that is, the weak conservation
equations now take the form, for $\ell\leq 6$,
\be\label{trijj}
\p_t \cCid{-\ell}{m_1\ldots m_{3\ell}} \sim \sum_{k \geq 1}
 \cpd{k}{n_1 \ldots n_{3k}}
 \cCid{-\ell -k}{m_1\ldots m_{3\ell}n_1\ldots n_{3k}},
\ee
with the {\em covariant} velocities $\cpt{k}{n_1 \ldots n_{3k}} (t)
\equiv e_{n_1}{}^{a_1} \cdots e_{n_{3k}}{}^{a_{3k}} \cpt{k}{a_1\dots a_{3k}}
= \p_t A_{n_1\dots n_{3k}} + \dots$. Index contractions here are only
schematic; we do not indicate the various (anti-)symmetri\-sations
required for the pertinent Young tableaux. The important feature of
(\ref{trijj}) is the distinction of contravariant and covariant world
indices.

This triangular evolution system can be recursively integrated, in the
present truncation starting from $\ell=6$. Indeed, the above procedure
eliminates all the terms on the r.h.s. of the last equation in
(\ref{timederconst}), implying that the contravariant constraint $\cCit{-6}{}$
is actually {\em constant}, and not only weakly constant. Due to the
identity (\ref{jandp}) between $J^{(-3)}$ and the contravariant
$\cpt{3}{}$ we can also rewrite $\cCit{-6}{}$ in
current$\times$current form and define
\be\label{l6}
\cCcd{-6}{m_1\ldots m_{10}||n_0|n_1\ldots n_7} \equiv
\cCid{-6}{m_1\ldots m_{10}||n_0|n_1\ldots n_7} =
  \cjd{-3}{n_0|m_1\ldots m_8}\cjd{-3}{m_9|m_{10}n_1\ldots n_7}\,.
\ee
Here, and in similar formulas below, the same symmetrisations as in
(\ref{constdef}) are understood. This way of writing the constraints
makes it plainly evident that $\cCct{-6}{}$ is strongly conserved since
the current components are. The notation ${\mathfrak{L}}$ is chosen in
anticipation of a Sugawara-like construction.

Examining then the weak conservation law for the penultimate (contravariant)
constraint $\cCit{-5}{}$, one finds that it, too, can be integrated
explicitly. More specifically, it is easy to check that the time
derivative of
\be
\cCcd{-5}{m_1\ldots m_{10}||n_1\ldots n_5}\equiv
 \cCid{-5}{m_1\ldots m_{10}||n_1\ldots n_5}
 +\frac1{15}A_{p_1p_2p_3}\cCid{-6}{m_1\ldots
 m_{10}||p_1|p_2p_3n_1\ldots n_5}
\ee
is identically zero by virtue of (\ref{timederconst5}) and (\ref{panda}).
After a little algebra, we find that, remarkably, we can again rewrite
this exactly conserved quantity in current $\times$ current form as
\be\label{l5}
\cCcd{-5}{m_1\ldots m_{10}||n_1\ldots n_5} &=&
   \cjd{-2}{n_1\ldots n_4m_1m_2}\cjd{-3}{n_5|m_3\ldots m_{10}}\,
\ee
by using (\ref{jandp}).

The recursive integration can be continued, in principle, for the other
constraints $\cCit{-4}{}$ and $\cCit{-3}{}$ and we anticipate that the so
obtained (`$\E$ covariantised') constraints $\cCct{-4}{}$ and $\cCct{-3}{}$
can also be expressed as bilinears in current components via
\bes{constdefnew}
\cCcd{-4}{m_1\ldots m_{10}||n_1n_2} &=&
   \frac{21}5 \cjd{-2}{n_1m_1\ldots m_5}\cjd{-2}{n_2m_6\ldots m_{10}}
  + \cjd{-1}{n_1m_1m_2}\cjd{-3}{n_2|m_3\ldots
    m_{10}}\,,\quad\quad\quad\mbox{ } \\
\cCcd{-3}{n_1\ldots n_9} &=&
   28 \cjd{-1}{n_1n_2n_3}\cjd{-2}{n_4\ldots n_9}
   + \cjd{0}{n_1}{}_p\cjd{-3}{p|n_2\ldots n_9}\,.
\ees
where we have substituted from (\ref{abvals}) for $\alpha$ and $\beta$.
We will show that these are the correct expressions below in
section~\ref{trmsec} by obtaining them from a symmetry transformation.
The above Sugawara-like, i.e. current $\times$ current, form of the
redefined constraints now renders manifest their strong
conservation.\footnote{In
  geometrical terms, strongly conserved (under geodesic motion)
  quantities which are linear in the velocities (such as $\cJ$) define
  `Killing vectors',  while strongly conserved quantities which are
  quadratic in the velocities define `Killing tensors'. It is a priori
  quite possible to have Killing tensors which cannot be expressed  in
  Sugawara form, i.e. as a combination of tensor products of Killing
  vectors (this is for instance the case for the Carter Killing tensor
  on a Kerr spacetime). Though we should leave open   the possible
  existence of such non-trivial Killing tensors at higher levels, it
  seems  that  `Sugawara-like', geometrically trivial Killing tensors
  are sufficient in our problem.}
The reason for switching to a labelling by negative levels for the
redefined constraints is now obvious: it follows immediately from the
level structure on the current components, and is such that
$\cCct{-\ell}{}= \sum_n  J^{(-\ell+n)} J^{(-n)}$, in a fashion very
similar to the Sugawara construction of the Virasoro generators for
affine algebras. This connection will be made more explicit in
section~\ref{affvir}. As is evident already from the few terms in
(\ref{constdefnew}) a Sugawara construction for $\E$ will be far
more intricate than the usual construction for affine algebras since
the tensorial structure on the various terms is different whereas in
the affine case only the level remains. Without truncation we also
expect formally infinite sums as extensions of
(\ref{constdefnew}).\footnote{In a quantum version these infinite sums
  presumably need to be normal ordered such that they become well-defined
  operators on any finite occupation number state.}

The above expressions (\ref{constdefnew}) at last furnish an
explanation why we need to introduce so many indices to parametrise
the constraints in (\ref{constdef}),
even though inspection of (\ref{sugraconst}) might suggest that a
more economical form of the constraints could be obtained by
dualizing and contracting out seemingly superfluous indices. Namely,
when written in contravariant form (\ref{constdefnew}), it is obvious
that we would need a metric $g_{mn}(t)$ (rather than merely a Kronecker
symbol $\delta_{ab}$) to contract away indices. However, the latter
metric is not a proper $\E$ object, and therefore a contracted version
of (\ref{constdefnew}) cannot possibly transform under $\E$ (or rather,
as we will see, $\E^+$) in the proper way; besides, contraction
with a time-dependent quantity would spoil strong conservation.

As is evident from the above construction, at the origin $\cV=1$
in coset space
the Sugawara-like constraints (\ref{constdefnew}) agree with
the weakly conserved ones in (\ref{constdef}), since the coset
zehnbein $e_m{}^a=\d_m{}^a$ and the additional coordinates
$A_{m_1m_2m_3}=A_{m_1\ldots m_6} =\ldots =0$.  Away from the origin
the identity $\cJ=\cP$ no longer holds and the constraints
(\ref{constdef}) and (\ref{constdefnew}) also start to differ. This is
captured by the contravariantisation by $e_m{}^a$, turning
$\cCt{\ell}{}$ into $\cCit{-\ell}{}$, and the additional terms proportional to
$A_{m_1m_2m_3}$ etc., turning $\cCit{-\ell}{}$ into the Sugawara-like
$\cCct{-\ell}{}$.

\end{subsection}

\begin{subsection}{Transformation of the constraints}
\label{trmsec}

We now examine the transformation properties of the constraints
under the basic $\E$ symmetry. This question can be addressed both for
the strongly conserved constraints (\ref{constdefnew}) quadratic in the
charges $\cJ$ as well as for the equivalent weakly conserved constraints
(\ref{constdef}) quadratic in the velocities $\cP$.

As already mentioned,
the tensor structure of the low level constraints is identical to that
of the so-called integrable $L(\Lambda_1)$  representation of $\E$.
The highest weight of this $\E$
representation is $\Lambda_1$ with
Dynkin labels $[1\,0\,0\,0\,0\,0\,0\,0\,0\,0]$, The `1'
here occurs for the over-extended, hyperbolic node of the $E_{10}$ Dynkin
diagram as shown in Fig.~\ref{e10dynk} and the definition of the
fundamental weights was given in footnote~\ref{funwt}.
The low level content of the $L(\Lambda_1)$ representation
(sometimes also referred to as `central charge representation') w.r.t.
the $A_9$ subgroup of $E_{10}$ was given in an appendix
of \cite{Kleinschmidt:2003jf}.\footnote{The analogue of
  $L(\Lambda_1)$ for $E_{11}$ was proposed in \cite{West:2003fc} to be
  responsible for the emergence of space-time. We will here take a
  different view on emergent space by the unfolding of constraint
  equations, but note that just like in (\ref{e11dec}) below further
  $E_{11}$ representations beyond $L(\Lambda_1)$ may be required there
  as well.}
Note however, that the constraint $\cCct{-6}{}$ contains only one
of the two $GL(10)$ tensors appearing in the level decomposition
of $L(\Lambda_1)$ due to the algebraic restriction that it should
vanish upon antisymmetrisation in the indices $n_0,n_1,\dots ,n_7$.
The possible occurrence of $L(\Lambda_1)$ of $\E$ in the present
context might be interpreted as evidence for a covariant
formulation involving $E_{11}$. In this case, gauge-fixing
and a canonical analysis should lead to the replacement of a (presently
unknown, and hypothetically $E_{11}$ invariant) set of `covariant'
equations by an $E_{10}$ invariant set of dynamical equations augmented
by constraint equations transforming in a representation of $E_{10}$,
whose structure should follow from an $E_{10}$ decomposition
of $E_{11}$. Indeed the $L(\Lambda_1)$ representation is the first in
an infinite sequence of integrable highest weight representations of
$E_{10}$ arising in the decomposition of $E_{11}$ w.r.t. its natural
$E_{10}$ subalgebra \cite{Kleinschmidt:2003pt}:
\be\label{e11dec}
E_{11} = \cdots  \oplus L(\Lambda_3)^* \oplus L(\Lambda_1)^*\oplus
(E_{10}\oplus \reals\kappa) \oplus L(\Lambda_1) \oplus L(\Lambda_3)
\oplus \cdots\,,
\ee
where we have also indicated the dual lowest weight representations
corresponding to the positive step operators (the fundamental weights
for $E_{11}$ are defined in a completely analogous fashion as for
$E_{10}$). $\kappa$ is a `level counting  operator' which commutes
with $E_{10}$ (and is analogous to
the central charge in the decomposition of $E_{10}$ under $E_9$).
If (\ref{e11dec}) were indeed the correct way of splitting the looked-for
$E_{11}$-covariant equations into dynamical $E_{10}$ equations and constraints,
one would accordingly expect an infinite set of (separately infinite)
towers of constraints, of which the known supergravity constraints
would just be the lowest lying members. The piece associated to
$\kappa$ is $E_{10}$ invariant and would correspond to the Hamiltonian
constraint.

In order to verify the $\E$ transformation properties of the
constraints it is better to work with the strongly conserved version
$\cCct{-\ell}{}$ since the constituent charges transform directly
under $\E$ according to (\ref{Jtrafo}). In contrast the weakly
conserved constraints $\cCt{\ell}{}$ of (\ref{constdef}) only transform
under the induced $\K$ transformation and we will study them below as
a second step.

The truncation condition (\ref{Jt}) is only maintained by the
parabolic subgroup $\E^+$. Let us start then by considering the
effect of an $\E^+$ transformation on the contravariantised constraints.
Using the transformation  (\ref{Jtrafo}) in
(\ref{l6}), (\ref{l5}) and (\ref{constdefnew}) we obtain under an
infinitesimal $\Lambda^{(1)}$ transformation of $\E$
\bes{newconstvar}
\d  \cCcd{-3}{m_1\ldots m_9} &=& -5
  \cld{1}{pqr}\cCcd{-4}{m_1\ldots m_9p||qr}\,,\\
\d   \cCcd{-4}{m_1\ldots m_{10}||n_1n_2} &=& \frac{5}{12}
  \cld{1}{pqr}\cCcd{-5}{m_1\ldots
    m_{10}||pqrn_1n_2}\,,\\
\d  \cCcd{-5}{m_1\ldots m_{10}||n_1\ldots n_5} &=&  \frac{1}{15}
   \cld{1}{pqr}\cCcd{-6}{m_1\ldots m_{10}||p|qrn_1\ldots
    n_5},\\
\d  \cCcd{-6}{m_1\ldots m_{10}||n_0|n_1\ldots n_7} &=& 0  \,,
\ees
where the last relation is again due to the truncation condition
(\ref{Jt}). The result (\ref{newconstvar}) exhibits two remarkable
features: $(i)$ the contravariantised constraints transform as a
linear representation (which was not at all guaranteed by their
definition), and $(ii)$ the linear transformations exhibited in
(\ref{newconstvar}) are the same as one would find for the
$L(\Lambda_1)$ representation of $\E$ restricted to $\E^+$.
We have already mentioned that $\mathfrak{L}_{-6}$ contains only
one of the two possible Young tableaux.
The set of constraints $\cCct{-3}{},\ldots,\cCct{-6}{}$
here furnish an unfaithful representation of $\E^+$ contained in
$L(\Lambda_1)$ of $\E$. However, it is not clear that this relation
between $L(\Lambda_1)$ and the constraints continues to hold when the
truncation is relaxed. We stress that the nice transformation laws
(\ref{newconstvar}) do {\em not} mean that the constraints constitute
a highest weight
representation of the full $\E$. In fact one can show that the
combination of $F_{[m_1m_2m_3}\otimes F_{m_4\ldots m_9]}$ and
$F_{n|[m_1\ldots m_8}\otimes K^n{}_{m_9]}$
of $\E$ generators underlying $\cCct{-3}{}$ is not
annihilated by the raising operator $E^{pqr}$. This suggests that the
tensor product of two adjoint representations $\cJ$ of $\E$ does not
contain $L(\L_1)$ as a subrepresentation. A similar result is known
in the affine case $E_9$
\cite{Chari:1987}.\footnote{\label{e9hstfn}However, in this case one can
  formally construct something like a highest weight vector.}

The difficulties with the transformation under full $\E$ also become
apparent when studying the way the weakly conserved constraints
(\ref{constdef}) change under the action of $\E$.
The infinitesimal
variation is determined by the induced $\K$ transformation, involving
both positive and negative step operators of $\E$. For simplicity we
restrict to the transformations (\ref{dp}) (with parameter
$\Lambda^{(1)}$);
the action of this transformation on the
constraint  $\cCt{3}{a_1\ldots  a_9}$ is
\be\label{constvar}
\d_{\clt{1}{}} \cCd{3}{a_1\ldots a_9} &=& -5
  \cld{1}{c_1c_2c_3}\cCd{4}{a_1\ldots a_9c_1||c_2c_3}
   -\frac12\cld{1}{a_1c_1c_2}\cpd{1}{c_1c_2c_3}\cpd{3}{c_3|a_2\ldots
     a_9}\nn\\
&& + 28\cld{1}{a_1a_2c}\cpd{0}{ca_3}\cpd{2}{a_4\ldots a_9}
   +56\cld{1}{a_1a_2a_3}\cpd{0}{ca_4}\cpd{2}{ca_5\ldots a_9}\,.
\ee
In general, to have covariance under the basic $\K$ transformation
$\d_{\Lambda^{(1)}} \cC^{(\ell)}$ would need to be equal to the sum of two
terms $\Lambda^{(1)} \cC^{(\ell+1)}$ and
$\Lambda^{(1)}\cC^{(\ell-1)}$. In the case $\ell=3$, as
$\cC^{(2)}$ does not (seem to) exist,
we would therefore like to have $\d_{\Lambda^{(1)}} \cC^{(3)}
\propto \Lambda^{(1)} \cC^{(4)}$.
The first term on the r.h.s. of (\ref{constvar}) ($\sim \L^{(1)}
\cC^{(4)}$) is thus the  expected  covariant transformation of constraint
into constraint and naturally agrees with the corresponding one in
(\ref{newconstvar}).   However, the other three terms
(containing $P^{(1)}P^{(3)}$ or $P^{(0)}P^{(2)}$)  do not rearrange into
combinations of constraints. It is not inconceivable that, when
relaxing the truncation to levels $ \ell \leq 3$ and considering a
non-zero value of $\cpt{4}{}$, the term containing
$P^{(1)}P^{(3)}$ might cancel against a term coming from the
variation of $\cpt{4}{}$ in a possible additional contribution $\sim
\cpt{1}{}\cpt{4}{}$ to the definition  of $\cCt{3}{}$.
However, this type of argument does not seem to apply for
the two last offending terms, of the type $\cpt{0}{}\cpt{2}{}$, in
(\ref{constvar}) which arise because of the $F_{pqr}$ piece in the
$\K$ transformation. Concerning the
latter problematic terms we note,
however, that they  arise because of the restricted  Young
symmetry   of $\cpt{3}{}$ --- if there was  an additional
anti-symmetric piece on   level $\ell=3$, equivalent to an additive
modification of $\cpt{3}{}$, these terms could superficially be made
to vanish. This remark could be indicative of a possible
Borcherds extension of $\E$. We will discuss this idea further in
the conclusions. Evidently, the constraints $\cC$ are invariant under
$g\in\E^+$ transformations since these do not induce any non-trivial
$\K$ transformations on $\cP$.

Even though the transformation properties are identical in either form
used for the constraints (strongly conserved or weakly conserved),
the geometrical status of the constraints is
somewhat clearer when considering them in Sugawara-like form
$\mathfrak{L}$. The
para\-bolic subgroup $\E^{+}$ has a transitive action on the coset $\E /
\K$ (this action is even essentially simply transitive, if we neglect
the minor ambiguity linked to the negative-root part of
$GL(10)$). Therefore we can use $\E^{+}$ as a group of `translations'
over the coset $\E / \K$. Similarly to the notion of `Clifford
translations' and `Clifford parallelism' in 3-dimensional elliptic
space, we can use $\E^{+}$ to translate, `in a parallel manner',  the
bundle of geodesics issued from the origin (i.e. the unit element)
to a different (and arbitrary) point in coset space. This symmetry
argument allows us to complete the proof (given in
section~\ref{sugsec} only for $\cCct{-6}{}$ and $\cCct{-5}{}$)
that the Sugawara-like $\cCct{-\ell}{}$ constraints are equivalent
to the $\cC^{(\ell)}$ ones. Indeed it suffices to parallelly
transport the $\cCct{-\ell}{}$ back to the origin where they agree
with the weakly conserved $\cC^{(\ell)}$.
The $\E^{+}$
covariance of the constraints means that this translation  operation
maps `good' (i.e. satisfying the constraints) geodesics stemming from
a point   to other good geodesics stemming from a different (and
arbitrary) point in coset space. On the other hand, the apparent
non-covariance  under the full $\E$ of the constraints means that the
set of good geodesics stemming from (say) the origin is not invariant
under the isotropy group leaving the origin fixed (which is the group
$\K$).
We will comment on this (unresolved) puzzle of partial loss of symmetry
in the concluding section.

 \end{subsection}

\begin{subsection}{Affine \bmath$E_9$\ubmath{} truncation and standard
    Sugawara construction}
\label{affvir}

We now show how the Sugawara-like form of the constraints
(\ref{constdefnew}) relates to the
well-known Sugawara construction of an associated Virasoro algebra
which exists for any affine Kac-Moody (current)
algebra  \cite{Sugawara:1967rw,Bardakci:1970nb,Goddard:1986bp}. This
is done by reducing to the affine $E_9\subset\E$.
If the Fourier modes of the (left or right) current $E_9$ are denoted
$j^a_n$ (where
$a$ is an $E_8$ Lie algebra index), the Fourier
modes of the associated Virasoro generators are of the form
$L_m \sim \sum_n j^a_{m-n}j^a_n$.

Let us consider the reduction of the $E_{10}$ conserved current $\cJ$,
and of the $E_{10}$ constraints, to $E_9$.
We discussed above the covariance of our $\E$ Sugawara construction under
`translations' by the transitive parabolic subgroup $\E^+$. We can exploit
this translation-covariance to limit ourselves to considering a null
geodesic starting (say at $t=0$) from the coset origin (i.e. the unit
 element of the group).
In this case we have (at $t=0$) $\cJ=\cP$ and $\cJ$ is therefore
symmetric.

The reduction of  $E_{10}$  to its natural affine subalgebra $E_9$
(as visible on their Dynkin diagrams fig.~\ref{e10dynk} by separating
out the first,
leftmost node labelled 1)
corresponds to letting all indices only range over $2,3,\ldots,10$. Then
the number of $2$s is related to the affine level in $E_9$, see
\cite{Kleinschmidt:2006dy} and below. In this
truncation all of the constraints (\ref{constdef}) identically vanish due to
the presence of the antisymmetric 10-tuples $[b_1\dots b_{10}]$, {\em except}
for the lowest one, $\cCt{3}{a_1\ldots a_9}$, which becomes an $SL(9)$
singlet. Using the notation of \cite{Kleinschmidt:2006dy} for the
$E_9$ current  algebra, we find
\footnote{The additional factor of
  $3$ for the level $3$ terms comes from changing to canonical
  normalisation of $E_8$. The central extension is $c=-K^1{}_1$ and
  the derivation $d=K^2{}_2$, see \cite{Kleinschmidt:2006dy} for
  details. Since we are working with the identity vielbein, the
  distinction between flat and curved $\E/\K$ indices is not necessary
  here.}
\be\label{Jexp}
\cJ &=& \frac3{9!} \cpd{3}{i|k_1\ldots k_8}\eps^{k_1\ldots k_8}
    (Z_{(0)}^i + Z_{i}^{(0)})
 + \frac{3\cdot 8}{9!} \cpd{3}{i|k_1\ldots k_7 2}
   \eps^{jk_1\ldots k_7} ( G_{(1)j}^i + G_{(-1)i}^j)\nn\\
&& +\frac{3\cdot 8}{9!} \cpd{3}{2|2k_1\ldots k_7}\eps^{k_1\ldots k_7i}
  (Z^{(2)}_i + Z_{(-2)}^i)
 + \frac1{6!}\cpd{2}{i_1\ldots i_6}
   (Z_{(0)}^{i_1\ldots i_6} + Z^{(0)}_{i_1\ldots i_6})\nn\\
&&+\frac1{3!\cdot 5!} \cpd{2}{2k_1\ldots k_5}\eps_{k_1\ldots k_5
   i_1i_2i_3}  (Z^{(1)}_{i_1i_2i_3} + Z_{(-1)}^{i_1i_2i_3})
  +\frac1{3!}\cpd{1}{i_1i_2i_3}
    (Z_{(0)}^{i_1i_2i_3} + Z_{i_1i_2i_3}^{(0)} )\nn\\
&&  + \frac1{2\cdot 6!}\cpd{1}{2k_1k_2}\eps_{k_1k_2i_1\ldots i_6}
    (Z^{(1)}_{i_1\ldots i_6} + Z_{(-1)}^{i_1\ldots i_6})
  + \frac12\cpd{0}{ij} (G_{(0)j}^i + G_{(0)i}^j)\nn\\
&&  + \cpd{0}{2i} (Z^i_{(-1)} + Z^{(1)}_i)
  + \left(\cpd{0}{22}+\cpd{0}{ii}\right) d
  - \left(\cpd{0}{11}+\cpd{0}{ii}\right) c\,.
\ee
The notation here is such that $i,j,\ldots$ are $SL(8)$ vector indices
and take
$3,4,\ldots, 10$; the index value $2$ corresponds to the affine $E_9$
extension of $E_8$, and at the same time labels the remaining
spatial coordinate $x^2$ in the dimensional reduction. The bracketed
sub- and superscripts on the $E_8$ generators $Z$ give the affine
level. $E_8$ itself is written in $SL(8)$ level decomposition as

\begin{center}
\begin{tabular}{ccccccccccccccc}
$E_8$  &=& ${\bf \bar{8}}$ &\op & ${\bf 28}$ &\op &${\bf \bar{56}}$ &\op
&$({\bf  63}\oplus {\bf 1})$ &\op &${\bf 56}$ &\op &${\bf \bar{28}}$&\op
&${\bf 8}$\\
&&$Z_i$ && $Z_{i_1\ldots i_6}$ && $Z_{i_1i_2i_3}$ && $G^i{}_j$ &&
  $Z^{i_1i_2i_3}$ && $Z^{i_1\ldots i_6}$ && $Z^i$
\end{tabular}
\end{center}
Using the current algebra basis, we expand the $E_9$ valued
conserved current (\ref{Jexp}) as
\be
\cJ &=& \sum_{m\in\ints}\big[J^{(m)}_i Z^{i}_{(m)}
  + J_{(m)}^i Z_{i}^{(m)}
  + \frac1{6!}J^{(m)}_{i_1\ldots i_6} Z^{i_1\ldots  i_6}_{(m)}
  + \frac1{6!}J_{(m)}^{i_1\ldots i_6} Z_{i_1\ldots  i_6}^{(m)}\nn\\
&& \quad + \frac1{3!}J^{(m)}_{i_1i_2i_3} Z^{i_1i_2i_3}_{(m)}
  + \frac1{3!}J_{(m)}^{i_1i_2i_3} Z_{i_1i_2i_3}^{(m)}
  + J^j_{(m)i} G^i_{(m)j} + J_d d + J_c c\big]\,.
\ee
In the present situation (\ref{Jexp}) only components up to
affine level $|m|\le 2$ are non-zero.

The na\"ive Sugawara construction\footnote{`Na\"ive' here means
  without normal ordering although this is superfluous anyway for the
  symmetric expansion used here. Ignoring normal ordering also makes
  the affine Sugawara generators invariant under the action of the
  affine group, in agreement with the observation in
  footnote~\ref{e9hstfn}.}
of the
Virasoro generator $L_{-1}$ gives with the standard normalisations
(in terms of charges and up to an overall factor)
\be\label{genL}
L_{-1} &=& J_i^{(-2)} J_{(1)}^i + J_i^{(-1)} J_{(0)}^i
 + J^i_{(-1)j} J^j_{(0)i} + J^i_{(-1)i} J^j_{(0)j}\nn\\
&& + \frac1{3!} J^{(-1)}_{i_1i_2i_3}J_{(0)}^{i_1i_2i_3}
 + \frac1{6!} J^{(-1)}_{i_1\ldots i_6}J_{(0)}^{i_1\ldots i_6}+\ldots\,,
\ee
where the dots vanish identically in the present truncation.

Substituting the above expansion (\ref{Jexp}) into  expression (\ref{genL})
gives
\be\label{lm1}
\frac13\eps_{k_1\ldots k_8}
L_{-1} &=& \frac1{9}\cpd{0}{2i}\cpd{3}{i|k_1\ldots k_8}
  + \frac{8}{9} \cpd{0}{2k_1}\cpd{3}{2|k_2\ldots
    k_82}
+\frac{8}{9} \cpd{0}{k_1i}\cpd{3}{i|k_2\ldots k_8
  2}\\
&& - \frac{8}{9} \cpd{0}{ii}\cpd{3}{2|k_1\ldots
  k_8}
+ 28\cdot\frac69 \cpd{2}{2k_1\ldots
    k_5}\cpd{1}{k_6k_7k_8}
 +28\cdot\frac39 \cpd{1}{2k_1k_2}\cpd{2}{k_3\ldots
  k_8}\,.\nn
\ee

This is to be compared with the reduction of $\cCct{-3}{}$ to $E_9$
which gives
\be\label{c3red}
\cCcd{-3}{2k_1\ldots k_8} &=& \frac19\cpd{0}{2i}\cpd{3}{i|k_1\ldots
  k_8} + \frac89 \cpd{0}{2k_1}\cpd{3}{2|k_2\ldots k_82}
+ \frac89 \cpd{0}{k_1i}\cpd{3}{i|k_2\ldots k_82}
+ \frac19\cpd{0}{22}\cpd{3}{2|k_1\ldots
  k_8} \nn\\
&& + 28\cdot\frac69 \cpd{1}{k_1k_2k_3}\cpd{2}{k_4\ldots k_8 2}
  + 28\cdot\frac39\cpd{1}{2k_1k_2}\cpd{2}{k_3\ldots k_8}\,.
\ee
where the normalisation is the same as in (\ref{constdef}). Remarkably, the
coefficients of most terms in (\ref{lm1}) and (\ref{c3red})
are identical, so that at least to the order considered, the momentum
constraint $\cCt{3}{}$ (generating translations along the residual
spatial coordinate $x^2$ \cite{Nicolai:1998gi}) appears
to be related to the $L_{-1}$ generator
(generating translations in the spectral parameter in a current algebra
realisation of $E_9$). The term which does not agree involves the
contribution $\cpt{2}{22}$ which in $\cJ$ of
(\ref{Jexp}) is only multiplied by $d$ and therefore cannot occur  in
$L_{-1}$.

This near-perfect agreement between the highest-level Sugawara-like
$E_{10}$ coset constraints (\ref{constdefnew}), and the
standard Sugawara construction of  $L_{-1}$ in the affine subalgebra $E_9$
suggests that the hyperbolic Sugawara-like definition of the
$E_{10}$ coset    constraints (obtained above only when assuming  a
truncation to levels $\leq 3$) should extend to the exact,
untruncated  coset model, and that the expected infinite tower of
$L(\L_1)$-like $E_{10}$ constraints  (of levels $-3, -4, -5,
\ldots$),  possibly with additional towers, should be   somehow
 analogous to the infinite tower of
Virasoro generators $L_m$.
In other words, the gauge symmetry of  M-theory in its $E_{10}$ formulation
would be contained in a vast generalization of the Sugawara construction
of the Virasoro constraints (encoding the conformal symmetry of a gauge-fixed
string action). Moreover, we expect that the supersymmetry constraint
(\ref{susycons}) arises as the part of the construction of the
fermionic current $G(z)$ as in string theory.

\end{subsection}

\end{section}

\begin{section}{Discussion}

In this concluding chapter we summarise our findings and outline some
interesting avenues of further research suggested by our results.

In this paper we have shown that it is possible to further constrain the
coset motion on $E_{10}/K(E_{10})$ by a set of weakly conserved constraints
(\ref{constdef}) which precisely correspond to the appropriate
truncations of the supergravity canonical constraints. We have also shown
that these constraints can be equivalently re-expressed in a Sugawara-like
manner (\ref{constdefnew}), which is interestingly linked to the standard
Sugawara construction of a Virasoro algebra for the natural affine
subalgebra $E_9$ of $E_{10}$. We have investigated these coset constraints
under the assumption of a (consistent) dynamical truncation where the
coset velocities $\cP$  (in triangular
gauge) of levels $4$ and higher, or more invariantly the conserved
charges $\cJ$ of levels $-4$ and below,  vanish. We conjecture that
if one relaxes this truncation by admitting non-zero coset charges down
to level $k$ it will be possible to extend both the definition of the
existing constraints (e.g. by adding new  terms, with total grading $-3$,
to $\cCct{-3}{}$ up to $ \cJ^{(-3+k)}\cJ^{(-k)}$, etc.), and
the number of constraints (by including lower-level constraints, down
to the level $\cCct{-2k}{}$). In the limit $k \to + \infty$
where the truncation is removed, one would end up with (at least) one
infinite tower of constraints $\cCct{-3 -n}{}$,
with $n \in \mathbb{N}$. Such an infinite
number of constraints might be conjectured to be needed to reduce the
potentially problematic  `exponentially infinite' number
of degrees of freedom in the hyperbolic $E_{10}/K(E_{10})$ coset model
to a more manageable size, hopefully compatible with the physically expected
M-theory degrees of freedom. On the other hand, this also means that the
set of constraints should not grow in size faster (when the level
increases) than the number of generators in $\E^+$.\footnote{From this
  point of view, one hopes that the apparent coincidence between the
  algebraic structure of the low-level constraints and the $\E$
  highest-weight representation $L(\L_1)$ does not persist to  all
  levels, because this representation grows in size faster than
  $\E^+$.}

However, it seems that the (say Sugawara-defined) coset constraints are
only covariant under a parabolic subgroup $\E^+$ of $\E$. The
constraint surface therefore seems to partially break the original
$E_{10}$ symmetry of the coset model. By contrast, let us remark that,
in the $E_9$-invariant (two-dimensional) reduction of supergravity
(which is closely related to the $E_9$ reduction of the $\E$ coset
model), $E_9$ {\em does} map solutions to solutions and so respects
the constraint surface. This point deserves further investigation, as
well as the relation between the hyperbolic and affine Sugawara constructions,
which might open a new perspective on the $E_{10}$
structure.

We see several possible `resolutions' of this partial loss of symmetry.
First, one might think of keeping the full $E_{10}$ Kac-Moody symmetry
by enlarging the set of constraints, i.e. by treating
the offensive terms in the transformation of the constraints
(\ref{constvar})  as additional constraints that need
to be imposed. Whereas we partially checked that this will lead to new
consistent conditions (of level $-2$ in Sugawara form)
on the geodesic motion, the transformation of
these new constraints again does not close covariantly but
necessitates yet again new terms (of level $-1$), etc..
 Anticipating that this phenomenon
persists indefinitely it is not clear to us whether any non-trivial
solutions to the geodesic equations remain in this process, given the
apparently very large, and maybe infinite\footnote{There is, however, the
possibility that, under our usual truncation, the above-defined set
of new constraints at levels $-2, -1, \ldots$ does terminate at level $+3$.
Actually, the fact that, at level $-6$, $\cCct{-6}{}$ seems already to
contain only one of the two independent corresponding
objects of $L(\Lambda_1)$ might be the first sign of such a
reduction of $L(\Lambda_1)$ to a smaller (possibly irreducible)
representation of $\E^+$.},
total set of constraints. In
addition, the new constraints do not appear to have a good
interpretation in supergravity.

A second possible resolution of the symmetry problem might
turn out to be  that $E_{10}$ is only an
auxiliary symmetry of the theory, which is broken by the
constraints. An analogy with, say, bosonic string theory, might
suggest how this could be the case. Indeed, the usual,
conformal-gauge-fixed dynamics of bosonic string theory consists of
two separate elements: {\em (i)} the conformal-gauge-fixed action
$S_{gf}[X^{\mu}(\tau,\sigma)]$, and {\em (ii)} the Virasoro constraints $L_n
=0$. The symmetry of $S_{gf}[X^{\mu}(\tau,\sigma)]$ include {\em both}
conformal transformations of worldsheet coordinates $(\tau,\sigma)$,
{\em and} the following (active) transformations of the target field
$X^{\mu}$: $\d X^{\mu} = \eps^{\mu}_n \exp (- i n (\tau \pm \sigma))$,
for arbitrary $n$ (and arbitrary choice of sign $\pm$ if one discusses
the closed string; for the open string one must combine terms of the
two signs). The Noether conserved current associated to the second
symmetry of $S_{gf}[X^{\mu}(\tau,\sigma)]$ is the worldsheet current $
j^{\mu}(\tau \pm \sigma)=  \p_{\pm} X^{\mu}$. The Fourier modes of
this Noether current are the usual $  j^{\mu}_n = \a^{\mu}_n$. The
current algebra is abelian (w.r.t. the `Lie algebra index' $\mu$) and
contains only the (level-1) anomaly term $ [j^{\mu}_m,j^{\mu}_n]= m
\eta^{\mu \nu} \d_{m+n}$.  The Virasoro constraints are obtained by
the standard Sugawara construction from the Noether current: $ L_m =
\frac{1}{2} \sum_n \eta^{\mu \nu} j^{\mu}_{m-n} j^{\nu}_n$. Now, the
crucial point we wish to make is that the constraints $ L_m =0$ are
{\em not covariant} under the current symmetry $  j^{\mu}_n =
\a^{\mu}_n$ of the gauge-fixed action
$S_{gf}[X^{\mu}(\tau,\sigma)]$. Indeed, the commutation relation
$[j^{\mu}_m,L_n]= m  j^{\mu}_{m+n}$ expresses the covariance of the
current under conformal transformations, but exhibits the
non-covariance of the constraints under the symmetry associated to
the current (only the Poincar\'e symmetry under the  zero-mode
$j^{\mu}_0$ is present). Let us also recall that the string
gauge-fixed Hamiltonian is proportional to the level-0 Virasoro
constraint $L_0$ (or $L_0 + \bar{L}_0$ in the closed string case), and
that this dissymetric role of $L_0$ w.r.t. the other $L_n$'s reflects
the specific gauge used to fix the worldsheet diffeomorphism
symmetry.  By analogy, it might happen that the
$E_{10}$  coset action is a gauge-fixed version of an underlying
gauge-invariant action, and that the gauge fixing has the effect of,
both, selecting one specific constraint ($\cC_0 = \langle \cP| \cP
\rangle = \langle \cJ| \cJ \rangle$) as Hamiltonian, and  introducing
an auxiliary symmetry which does not preserve the constraints. We
might, however, expect that  (as in the string case where the symmetry
generators, $  j^{\mu}_n = \a^{\mu}_n$ generate the full spectrum) the
auxiliary symmetry be a spectrum generating symmetry. And in the case
where the analogy should be taken with the light-cone-gauge-fixed
string action, the auxiliary symmetry might be similar to a DDF algebra.

An alternative explanation of the apparent restriction to $\E^+$, one
might need to modify the underlying $E_{10}$ Kac-Moody symmetry
(e.g. into a related Borcherds symmetry, see previous subsection).
We finally come back to the possibility of a Borcherds extension
mentioned in section~\ref{trmsec} in connection with the `missing'
full $\E$ covariance of the constraint transformations.
Supposing that this implies that  $E_{10}$ may not
be the correct full symmetry structure of the complete supergravity
system (and of M-theory) one should look for a modification of $\E$
preserving the remarkable features of the $\E$ model. Arguably the
simplest modification of $E_{10}$
is an extension of $E_{10}$ by additional simple
generators.
Adjoining new {\em imaginary simple roots} leads to a {\em Borcherds
extension} of $\E$. (For introductory literature to Borcherds
algebras see for
example \cite{Borcherds}.)  Introducing a new anti-symmetric nine index
generator corresponding to salvaging the transformation properties
of (\ref{constvar}) can be achieved by means of adding a single null
imaginary simple root which attaches with a single line at the
hyperbolic node of the $E_{10}$ Dynkin diagram. (In supergravity
terms such a new generator  relates to the spatial trace of the spin
connection.)
However, the
transformation of the associated component of $\cP$ under an
infinitesimal $\L_{a_1a_2a_3}$
transformation as in (\ref{dp}) does not give rise (as would be needed
to cancel the offending terms $\propto P^{(0)}P^{(2)}$) to a new
contribution to $\d\cpt{3}{}$ in (\ref{varp3}) precisely since the new
root is simple and not composite. Therefore such a new root does not
appear to correct the transformation properties of the constraints.
Another possible Borcherds extension is the vertex
operator  algebra obtained from a lattice construction on the $E_{10}$
root lattice (see for example \cite{Gebert:1994mv}). This Borcherds
algebra also contains $E_{10}$ as a proper subalgebra but has
additional imaginary simple roots. The first such imaginary simple
root is time-like and as a root vector identical to $\Lambda_2$ and
therefore occurs at level $\ell=6$, much too high a level than
needed to correct the  constraint transformation (\ref{constvar}).
We note, however, that a Borcherds modification of $E_{10}$ (maybe
involving several independent copies of the null imaginary root
mentioned above) could also help to produce additional negative
definite terms in the Hamiltonian constraint
$\langle \cP|\cP\rangle=0$ in order to improve agreement with
supergravity \cite{Damour:2004zy}.

To further investigate the situation, it will be important to compute the
algebra generated by the constraints. This should give access
to the underlying gauge symmetry of the full model.  From the supergravity
correspondence (proven at low levels only), we expect that
the bosonic coset constraints will contain a generalization of both
diffeomorphism invariance, and the gauge invariance of the 3-form.
It would be interesting to see whether a richer algebraic
structure, maybe appropriate to a theory in which both spacetime and its
general covariance are expected to be emergent properties,
comes out of a  group-theoretical analysis of the constraint algebra.

\end{section}

\vspace*{1cm}

\noindent {\bf Acknowledgements}\\
\indent AK and HN gratefully acknowledge the hospitality of IHES during
several visits, while TD parallelly acknowledges the recurrent hospitality of the AEI.
This work was partly supported by the European Research and
Training Networks `Superstrings' (contract number MRTN-CT-2004-512194)
and `Forces Universe' (contract number MRTN-CT-2004-005104).


\begin{thebibliography}{10}

\bibitem{Damour:2002cu}
  T.~Damour, M.~Henneaux and H.~Nicolai,
  Phys.\ Rev.\ Lett.\  {\bf 89} (2002) 221601
  [arXiv:hep-th/0207267].

\bibitem{Kleinschmidt:2004dy}
  A.~Kleinschmidt and H.~Nicolai,
  JHEP {\bf 0407} (2004) 041
  [arXiv:hep-th/0407101].

\bibitem{Damour:2004zy}
  T.~Damour and H.~Nicolai,
  arXiv:hep-th/0410245.

\bibitem{Kleinschmidt:2004rg}
  A.~Kleinschmidt and H.~Nicolai,
  Phys.\ Lett.\  B {\bf 606} (2005) 391
  [arXiv:hep-th/0411225].

\bibitem{Damour:2005zs}
  T.~Damour, A.~Kleinschmidt and H.~Nicolai,
  Phys.\ Lett.\  B {\bf 634} (2006) 319
  [arXiv:hep-th/0512163].

\bibitem{de Buyl:2005mt}
  S.~de Buyl, M.~Henneaux and L.~Paulot,
  JHEP {\bf 0602} (2006) 056
  [arXiv:hep-th/0512292].

\bibitem{Kleinschmidt:2006tm}
  A.~Kleinschmidt and H.~Nicolai,
  Phys.\ Lett.\  B {\bf 637} (2006) 107
  [arXiv:hep-th/0603205].

\bibitem{Damour:2006xu}
  T.~Damour, A.~Kleinschmidt and H.~Nicolai,
  JHEP {\bf 0608} (2006) 046
  [arXiv:hep-th/0606105].

\bibitem{West:2000ga}
  P.~C.~West,
  JHEP {\bf 0008} (2000) 007
  [arXiv:hep-th/0005270].

\bibitem{West:2001as}
  P.~C.~West,
  Class.\ Quant.\ Grav.\  {\bf 18}, 4443 (2001)
  [arXiv:hep-th/0104081].

\bibitem{West:2003fc}
  P.~C.~West,
  Phys.\ Lett.\  B {\bf 575} (2003) 333
  [arXiv:hep-th/0307098].

\bibitem{Damour:2000wm}
  T.~Damour and M.~Henneaux,
  Phys.\ Rev.\ Lett.\  {\bf 85} (2000) 920
  [arXiv:hep-th/0003139].

\bibitem{Damour:2000hv}
  T.~Damour and M.~Henneaux,
  Phys.\ Rev.\ Lett.\  {\bf 86} (2001) 4749
  [arXiv:hep-th/0012172].

\bibitem{Damour:2002et}
  T.~Damour, M.~Henneaux and H.~Nicolai,
  Class.\ Quant.\ Grav.\  {\bf 20} (2003) R145
  [arXiv:hep-th/0212256].

\bibitem{Damour:2005zb}
  T.~Damour and H.~Nicolai,
  Class.\ Quant.\ Grav.\  {\bf 22} (2005) 2849
  [arXiv:hep-th/0504153].

\bibitem{DHHS} J.~Demaret, J.L.~Hanquin, M.~Henneaux and P.~Spindel,
    Nucl. Phys. {\bf B252} (1985) 538

\bibitem{CJS} E.~Cremmer, B.~Julia and J.~Scherk,
  Phys. Lett. B {\bf 76} (1978) 409--412

\bibitem{Hillmann:2006ic}
  C.~Hillmann and A.~Kleinschmidt,
  Gen.\ Rel.\ Grav.\  {\bf 38} (2006) 1861
  [arXiv:hep-th/0608092].

\bibitem{Kac} V.G.~Kac, Infinite Dimensional Lie Algebras, 3rd edn.,
 Cambridge University Press, 1990

\bibitem{Thiemann:2001yy}
  T.~Thiemann,
  arXiv:gr-qc/0110034.

\bibitem{Nicolai:2005mc}
  H.~Nicolai, K.~Peeters and M.~Zamaklar,
  Class.\ Quant.\ Grav.\  {\bf 22} (2005) R193
  [arXiv:hep-th/0501114].

\bibitem{DN}
 T.~Damour and H.~Nicolai,
 arXiv:0705.2643[hep-th].

\bibitem{Sugawara:1967rw}
  H.~Sugawara,
  Phys.\ Rev.\  {\bf 170} (1968) 1659.

\bibitem{Bardakci:1970nb}
  K.~Bardak\c{c}i and M.~B.~Halpern,
  Phys.\ Rev.\  D {\bf 3} (1971) 2493.

\bibitem{Goddard:1986bp}
  P.~Goddard and D.~I.~Olive,
  Int.\ J.\ Mod.\ Phys.\  A {\bf 1} (1986) 303.

\bibitem{BM} P.~Breitenlohner and D.~Maison,
  Ann.\ Inst.\ H.~Poincar\'e, Phys. Th\'eor. {\bf 46} (1987) 215

\bibitem{Julia:1996nu}
  B.~Julia and H.~Nicolai,
  Nucl.\ Phys.\  B {\bf 482} (1996) 431
  [arXiv:hep-th/9608082].

\bibitem{Kleinschmidt:2003jf}
  A.~Kleinschmidt and P.~C.~West,
  JHEP {\bf 0402} (2004) 033
  [arXiv:hep-th/0312247].

\bibitem{de Buyl:2005zy}
  S.~de Buyl, M.~Henneaux and L.~Paulot,
  Class.\ Quant.\ Grav.\  {\bf 22} (2005) 3595
  [arXiv:hep-th/0506009].

\bibitem{Kleinschmidt:2003pt}
  A.~Kleinschmidt,
  Nucl.\ Phys.\  B {\bf 677} (2004) 553
  [arXiv:hep-th/0304246].

\bibitem{Kleinschmidt:2006dy}
  A.~Kleinschmidt, H.~Nicolai and J.~Palmkvist,
  JHEP {\bf 0706} (2007) 051
  [hep-th/0611314].

\bibitem{Nicolai:1998gi}
  H.~Nicolai and H.~Samtleben,
  Nucl.\ Phys.\  B {\bf 533} (1998) 210
  [arXiv:hep-th/9804152].

\bibitem{Chari:1987} V.~Chari and A.~Pressley,
  Math. Ann. {\bf 277} (1987) 543--562

\bibitem{Borcherds} R.~E.~Borcherds,
   J. Algebra {\bf 115} (1988) 501--512;
   J. Algebra {\bf 174} (1995)  1073--1079.
  E.~Jurisich,
   Contemp. Math. {\bf 194} (1996) 121--159

\bibitem{Gebert:1994mv}
  R.~W.~Gebert and H.~Nicolai,
  Commun.\ Math.\ Phys.\  {\bf 172} (1995) 571
  [arXiv:hep-th/9406175].

\end{thebibliography}
\end{document}